\documentclass[12pt,letterpaper]{article}
\RequirePackage[colorlinks,citecolor=blue,urlcolor=blue]{hyperref}
\usepackage[utf8]{inputenc} 

\usepackage{geometry} 
\usepackage{graphicx} 

\usepackage{booktabs} 
\usepackage{array} 
\usepackage{paralist} 
\usepackage{verbatim} 
\usepackage{amsbsy}
\usepackage{amsmath}
\usepackage{longtable}
\usepackage[nolists]{endfloat}
\usepackage[toc,page]{appendix}
\usepackage{natbib}
\usepackage{url}

\usepackage{breakurl}

\usepackage{amssymb}

\usepackage[textsize=small]{todonotes}
\usepackage{subcaption}

\usepackage{multirow}

\usepackage{dcolumn}

\raggedright
\graphicspath{{../figures/}}

\def\facmat{\mathbf{\facmatsm}}
\def\facmatsm{F}
\def\loadmat{\mathbf{\loadmatsm}}
\def\loadmatsm{L}
\def\datamatsm{Y}
\def\datamat{\mathbf{\datamatsm}}
\def\workmatsm{Z}
\def\workmat{\mathbf{\workmatsm}}
\def\rowvarresidsm{\Lambda}
\def\rowvarresid{\mathbf{\Lambda}}

\def\loadmean{\mu}

\def\resid{\boldsymbol{\epsilon}}

\def\cutpoint{\gamma}

\def\sequence{\mathbf{S}}

\def\treeparam{{\cal{F}}}
\def\treeparamsm{\tau}
\def\branchrates{\mathbf{B}}

\def\facprecTree{\boldsymbol{\Psi}_{\treeparam}}

\def\rowvarresidshape{\alpha_{\rowvarresid}}
\def\rowvarresidrate{\beta_{\rowvarresid}}

\def\loadprec{\lambda}

\def\nfac{K}
\def\ntaxa{N}
\def\ntraits{P}

\def\facprecDiff{\mathbf{I}_{\nfac}}
\def\colvarresid{\mathbf{I}_{\ntaxa}}

\def\ida{i}
\def\idb{j}
\def\idaa{i'}

\def\idc{k}

\def\notraits{m}

\def\pathParameter{\beta}
\def\paramList{\boldsymbol{\theta}}

\def\treeTraitsRoot{\boldsymbol{\mu}_R}

\def\diffprecprior{\boldsymbol{\Lambda}_{R_0}}
\def\diffvar{\boldsymbol{\Sigma}}
\def\diffmean{\boldsymbol{\mu}_0}
\def\pss{\kappa_0}
\def\diffprecpriordf{\nu}

\newcommand{\cdensity}[2]{\ensuremath{p(#1 \,|\,#2)}}  
\newcommand{\density}[1]{\ensuremath{p(#1 )}}

\newcommand{\trOperator}[1]{\mbox{tr}\left[ #1 \right]}
\newcommand{\oneVector}{\mathbf{1}}
\newcommand{\order}[1]{{\cal O} \left( #1 \right)}

\renewcommand{\section}[1]{%
\begin{center}
\begin{Large}
\normalfont\scshape #1
\medskip
\end{Large}
\end{center}}

\renewcommand{\subsection}[1]{%
\begin{center}
\begin{large}
\normalfont\itshape #1
\medskip
\end{large}
\end{center}}

\renewcommand{\paragraph}[1]{%
\vspace{2ex}
\noindent
\textit{#1.}---}

\renewcommand{\tableofcontents}{}

\bibpunct{(}{)}{;}{a}{}{,}  

\begin{document}
\begin{flushright}
Version dated: \today
\end{flushright}

\bigskip
\medskip
\begin{center}

\noindent{\Large \bf Phylogenetic Factor Analysis}
\bigskip

\noindent{\normalsize \sc
	Max R.~Tolkoff$^1$,
	Michael E.~Alfaro$^{2}$,
	Guy Baele$^{3}$,
	Philippe Lemey$^{3}$, \\
    and Marc A.~Suchard$^{1,4,5}$}\\
\noindent {\small
  \it $^1$Department of Biostatistics, Jonathan and Karin Fielding School of Public Health, University
  of California, Los Angeles, United States} \\
  \it $^2$Department of Ecology and Evolutionary Biology, University of California, Los Angeles, United States \\
  \it $^3$Department of Microbiology and Immunology, Rega Institute, KU Leuven, Leuven, Belgium \\
  \it $^4$Department of Biomathematics, David Geffen School of Medicine at UCLA, University of California,
  Los Angeles, United States \\
  \it $^5$Department of Human Genetics, David Geffen School of Medicine at UCLA, Universtiy of California,
  Los Angeles, United States

\end{center}
\medskip
\noindent{\bf Corresponding author:} Marc A. Suchard, Departments of Biostatistics, Biomathematics, and Human Genetics, 
University of California, Los Angeles, 695 Charles E. Young Dr., South,
Los Angeles, CA 90095-7088, USA; E-mail: \url{msuchard@ucla.edu}\\

\vspace{1in}

\clearpage

\paragraph{Abstract}

Phylogenetic comparative methods explore the relationships between quantitative traits adjusting for shared evolutionary history.
This adjustment often occurs through a Brownian diffusion process along the branches of the phylogeny that generates model residuals or the traits themselves.
For high-dimensional traits, inferring all pair-wise correlations within the multivariate diffusion is limiting.
To circumvent this problem, we propose phylogenetic factor analysis (PFA) that assumes a small unknown number of independent evolutionary factors arise along the phylogeny and these factors generate clusters of dependent traits.
Set in a Bayesian framework, PFA provides measures of uncertainty on the factor number and groupings, combines both continuous and discrete traits, integrates over missing measurements and incorporates phylogenetic uncertainty with the help of molecular sequences.
We develop Gibbs samplers based on dynamic programming to estimate the PFA posterior distribution, 
over three-fold faster 
than for multivariate diffusion and a further order-of-magnitude more efficiently
in the presence of latent traits.
We further propose a novel marginal likelihood estimator for previously impractical models with discrete data and find that PFA also provides a better fit than multivariate diffusion
in evolutionary questions in columbine flower development, placental reproduction transitions and triggerfish fin morphometry.

\noindent (Keywords: Bayesian inference; comparative methods; morphometrics; phylogenetics)\\

\clearpage

\label{ch:PFA}
\section{Introduction}

Phylogenetic comparative methods revolve around uncovering relationships between different characteristics or traits of a set of organisms over the course of their evolution. 
One way to gain insight into these interactions is to analyze unadjusted correlations between traits across taxa. 
However, as insightfully noted by \cite{Felsenstein85}, unadjusted analyses introduce the inherent challenge that any association uncovered may reflect the shared evolutionary history of the organisms being studied, and hence their similar traits values, rather than processes driving traits to co-vary over time. 
Thus, studies to identify co-varying evolutionary trait processes must simultaneously adjust for shared evolutionary history. 

There have been many attempts to accomplish this goal. 
\cite{Felsenstein85} and \cite{phylogenetic_least_squares} are two such important examples, but they rely on a known evolutionary history described by a fixed phylogenetic tree and consider univariate evolutionary processes giving rise to only single traits.
\cite{Felsenstein85} treats continuous traits as undergoing conditionally independent, Brownian diffusion down the branches of the phylogenetic tree and \cite{phylogenetic_least_squares} posit a regression model where the tree determines the error structure in the univariate outcome model.
\cite{HuelsenbeckRannala2003} adapt the Brownian diffusion description in a Bayesian framework
with the goal of drawing simultaneous inference on both the tree from molecular sequence data as well as the correlations of interest related to a small number of traits through a multivariate Brownian diffusion process. 
\cite{brownian_diffusion_phylogeography} extend the multivariate process by relaxing the strict Brownian assumption along distinct branches in the tree using a scale mixture of normals representation.
\cite{cybis2015assessing} jointly model molecular sequence data and multiple traits using a multivariate latent liability formulation to combine both continuous and discrete observations and determine their correlation structure while adjusting for shared ancestry.
This method is effective, but inference remains computationally expensive and estimates of the high-dimensional correlation matrix between latent traits is often difficult to interpret when addressing scientifically relevant questions.
Additional frequentist methods include, \cite{revell2009size} who use a phylogenetically adjusted principal components analysis, \cite{Adams14PLGS} who use a phylogenetic least squares analysis, and \cite{Clavel15Multivariate} who also use a multivariate diffusion method. 
All of these methods, however require large matrix inversions which make them ill suited to adaptations to full Bayesian inference, or bootstrapping to provide measures of uncertainty.

\par

One way to alleviate these problems lies with dimension reduction through exploratory factor analysis \citep{Aguilar00bayesiandynamic}. 
Factor analysis is the inferred decomposition of observed data into two matrices, a factor matrix representing a set of underlying unobserved characteristics of the subject which give rise to the observed characteristics and a loadings matrix which explains the relationship between the unobserved and observed characteristics.
Another form of dimension reduction through matrix decomposition is an eigen decomposition known as a principal components analysis (PCA).
\cite{phylogenetic_factor_analysis_masters} provides a method for constructing PCA adjusted for evolutionary history. 
This method, however, has the same problems typically associated with PCA, namely that it is not invariant to the scaling of the data and the elimination of the smaller components necessitates some information loss. 
In a frequentist setting, the author also provides no approach for simultaneous inference on the phylogenetic tree that is rarely known without error \citep{HuelsenbeckRannala2003}.
In addition, there lacks a reasonable prescription for measuring  
uncertainty about which traits contribute to which principle components.
\cite{FA_tree} design a factor analysis method which uses a Kingman coalescent to construct a dendrogram across a factor analysis for genetic data. 
While this is similar to the idea we will employ, this specific method uses a dendrogram between, rather than within, factors and is thus ill suited to handle the 
important problem we tackle in this paper.
Namely, researchers often seek to identify a small number of relatively independent evolutionary processes, each represented by a factor changing over the tree, that ultimately give rise to a large number of observed, dependent traits.

To formulate such a phylogenetic factor analysis (PFA) model, we begin with usual Bayesian factor analysis, as posited by \cite{LopesWest} and \cite{IFAPolisci}, which represents underlying latent characteristics of a group of organisms through a factor matrix and maps those latent characteristics to observed characteristics via a loadings matrix.
In a standard factor analysis, the underlying factors for each species would be assumed to be independent of each other, however this does nothing to adjust for evolutionary history.
\cite{Vrancken2015} describe how a high-dimensional Brownian diffusion can be used to describe the relationship between all of these observed traits, however the signal strength of the results of analyzing this model can be quite poor.
By using independent Brownian diffusion priors on our factors, our PFA model groups traits into a parsimonious number of factors while successfully adjusting for phylogeny.
Scientifically, these diffusions represent independent evolutionary processes.
We use Markov chain Monte Carlo (MCMC) 
integration in order to draw inference on our model through a Metropolis-within-Gibbs approach.
This facilitates both a latent data representation \citep{cybis2015assessing} for integrating discrete and continuous traits and a natural method to handle missing data relevant to our problems.
We further rely on path sampling methods \citep{PathSampling} to determine the appropriate number of factors \citep{Ghosh09defaultprior}. Since the latent, probit model necessitates the use of hard thresholds, we now have introduced an inherent difficulty in path sampling. 
In order to get around this difficulty, we employ a novel method which relies on softening the threshold necessitated by the probit model slowly over the course of the path. 
We additionally develop a novel method by which to handle identifiability issues inherent to factor analysis by taking advantage of the fact that correlated elements in the loadings matrix tend to be correlated across the MCMC chain. 

We show that our PFA method performs superiorly to a high-dimensional Brownian diffusion in both signal strength and, when we are inferring large numbers of latent traits, speed using the examples of the evolution of the flower genus  \textit{Aquilegia}, as well as the reproduction of the fish family \textit{Poeciliidae} that involves trait 
measurements missing at random.
Lastly, we explore the dorsal, anal and pectoral fin shapes of the fish family \textit{Balistidae} in order to explore this method's ability to handle situations where the number of traits are large compared to the number of species and to explore the simultaneous inference on our method along with the evolutionary history of these organisms with the aid of sequence data. The PFA model and its inference tools will be released in the popular phylogenetic inference package BEAST \citep{BEAST}.

\section{Methods}

\subsection{Phenotypic Trait Evolution}


%
%
Consider a collection of $\ntaxa$ biological entities (taxa). 
From each taxon \(\ida=1,\ldots,\ntaxa\), we observe a \(\ntraits\)-dimensional measurement \(\datamat_\ida=(\datamatsm_{\ida 1},\ldots,\datamatsm_{\ida \ntraits})\) of traits and, if available, a molecular sequence $\sequence_{\ida}$.
We organize these phenotypic traits into an \(\ntaxa \times \ntraits\) matrix \(\datamat = (\datamat_1,\ldots,\datamat_\ntaxa)'\) and an aligned sequence matrix $\sequence$. 
These taxa are related to each other through an evolutionary history \(\treeparam\), informed through $\sequence$, and we are interested in learning about the evolutionary processes along this history that give rise to observed traits $\datamat$. 


The history \(\treeparam\) consists of a tree topology \(\treeparamsm\) and a series of branch lengths \(\branchrates\). 
The tree topology is a bifurcating directed acyclic graph with a single generating point called the root, representing the most recent common ancestor of the given taxa, and with end points, each of which corresponds to a different taxon. 
The branch lengths correspond to edge weights of the graph, reflecting the evolutionary time before bifurcations. 
The history \(\treeparam\) may be known and fixed, or unknown and jointly inferred using $\datamat$ and $\sequence$. 
For further details on constructing the sequence-informed prior distribution $p( \treeparam \, | \, \sequence)$ and integrating over $\treeparam$ when unknown, see, e.g., \citet{marc2001} or \citet{BEAST}. 

In order to simultaneously model continuous, binary and ordinal traits, we adapt a latent data representation through the partially observed, standardized matrix 
$\workmat$
with entries
\begin{equation}
\workmatsm_{\ida \idb}
= \left\{
\begin{array}{l l}
(\datamatsm_{\ida\idb} - \hat{\datamatsm}_{\idb}) / \hat{\sigma}_{\idb}  & \text{ if trait } \idb \text{ is continuous }\\
\workmatsm_{\ida\idb} & \text{ if trait } \idb \text{ is binary or ordinal, }
\end{array}
\right.
\end{equation}
where 
$\hat{\datamatsm}_{\idb}$ is the mean of trait $\idb$ across taxa,
$\hat{\sigma}_{\idb}$ is its standard deviation for $\idb=1,\ldots,\ntraits$ and, more importantly,
$\workmatsm_{\ida\idb} \in \mathbb{R}$ is an unknown random variable that satisfies the restrictions
\begin{equation}
\cutpoint_{\idb(c-1)} < \workmatsm_{\ida\idb} \le \cutpoint_{\idb c} \text{ given } \datamatsm_{\ida\idb}=c
\label{eq:restriction}
\end{equation}
and 
$c \in \{1,\ldots,\notraits_\idb\}$ for $\notraits_\idb$-valued binary/ordinal data for trait $\idb$.
\newcommand{\allCutpoint}{\boldsymbol{\cutpoint}}
%
For identifiability, latent trait cut-points $\allCutpoint_\idb=(\cutpoint_{\idb 0}, \dots, \cutpoint_{\idb \notraits_{\idb}})$
take on the restrictions 
$\cutpoint_{\idb 0}=-\infty$, 
$\cutpoint_{\idb 1}=0$ and 
$\cutpoint_{\idb \notraits_\idb}=\infty$ 
or are otherwise random and jointly inferred.
Grouping cut-points for all binary or ordinal traits into $\allCutpoint$, \citet{cybis2015assessing} suggest assuming that differences between the small number of successive, random cut-points are \textit{a priori} exponentially distributed with mean $\frac{1}{2}$ to define their density $\density{\allCutpoint}$.  
\cite{cybis2015assessing} also discuss in detail how to treat categorical data in this sort of analysis. Since we do not use examples which contain non-ordered categorical data we elect not to describe those methods in these sections, but we will mention that they are implemented in BEAST and are easily adapted to fit the methods described in this paper.

In order to uncover the biological relationships amongst traits in $\workmat$ while controlling for evolutionary history, previous work relies on a Gaussian process generative model induced through considering conditionally independent Brownian diffusion along each branch in \(\treeparam\) \citep{Felsenstein85}. 
In a multivariate setting, a $\ntraits \times \ntraits$ 
variance matrix $\diffvar$ 
and unobserved, $\ntraits$-dimensional root trait value $\treeTraitsRoot$ characterize the process.
\citet{pybus2012} identify that analytic integration of $\treeTraitsRoot$ is possible by assuming that $\treeTraitsRoot$ is \textit{a priori} multivariate normally distributed with a fixed hyperprior mean $\diffmean$ and variance equal to $\pss^{-1} \diffvar$, where $\pss$ is a fixed hyperprior sample-size.
Consequentially, given $\treeparam$ and $\diffvar$, the latent traits $\workmat$ are distributed according to a matrix-normal (MN)
\newcommand{\minusSpace}{-0.2em}
\newcommand{\mn}[3]{\text{MN}\hspace{\minusSpace}\left( #1, #2, #3 \right)}
\newcommand{\mvn}[2]{\text{MVN}\hspace{\minusSpace}\left( #1, #2 \right)}
\newcommand{\transpose}{^{t}}
\begin{equation}
\workmat \sim \mn{
\diffmean}{ 
\facprecTree + \pss^{-1} \mathbf{J}}{
\diffvar},
\label{eqn:fullDiff}
\end{equation}
where
$\facprecTree  + \pss^{-1} \mathbf{J}$ is the across-taxa (row) variance and a deterministic function of phylogeny $\treeparam$,
$\diffvar$ is the across-trait (column) variance,
and
\(\mathbf{J}\) is a $\ntaxa \times \ntaxa$ matrix of ones \citep{Vrancken2015}.
Traits $\workmat$ have density function
\begin{equation}
\begin{aligned}
\label{eq:matrix-normal}
\cdensity{\workmat}{\treeparam, \diffvar}
=
\frac{
\mbox{exp}
\left\{
	-\frac{1}{2} \trOperator{
		\diffvar^{-1}
		\left( \workmat - \oneVector\diffmean\transpose \right)\transpose		
		\left( \facprecTree + \pss^{-1} \mathbf{J} \right)^{-1} 
		\left( \workmat - \oneVector\diffmean\transpose \right)		
}
\right\}
}{
	\left(2 \pi\right)^{\ntaxa\ntraits/2}	
	\left| \diffvar \right| ^{\ntaxa/2}
	\left| \facprecTree + \pss^{-1} \mathbf{J} \right|^{\ntraits/2}
}
,
\end{aligned}
\end{equation}
where $\trOperator{\cdot}$ is the trace operator and $\oneVector$ is a $\ntaxa$-dimensional column vector of ones.
Tree variance matrix \(\facprecTree\) contains diagonal elements that are equal to the sum of the adjusted branch lengths in \(\treeparam\) between the root node and taxon \(\ida\), and off-diagonal elements \((\ida,\idaa)\) that are equal to the sum of the adjusted branch lengths between the root node and the most recent common ancestor of taxa \(\ida\) and \(\idaa\), where the adjusted branch lengths represent a function of wall time and a branch rate accounting for variation in evolution rate over the course of the tree. For our diffusion model, we scale our tree such that from the root to the most recent tip we say that the process has undergone one diffusion unit. 

\par

Placing a conjugate prior distribution on $\diffvar$, such as 
$\diffvar^{-1}\sim \text{Wishart}_\diffprecpriordf(\diffprecprior)$
where \(\diffprecpriordf\) is the hyperprior degrees of freedom and \(\diffprecprior\) is the hyperprior belief on the structure of the inverse of the variance matrix \(\diffvar\), enables inference about its posterior distribution, shedding light on how the evolution of these traits relate to each other.
Such inference often requires repeated evaluation of density (\ref{eq:matrix-normal}), especially when the phylogeny $\treeparam$ or variance $\diffvar$ is random.
This evaluation suggests a computational order $\order{\ntaxa^3 + \ntraits^3}$, arising from the inversion of the $\ntaxa \times \ntaxa$ variance matrix $\facprecTree + \pss^{-1} \mathbf{J}$ and $\ntraits \times \ntraits$ variance matrix $\diffvar$.
One easily avoids the latter by parameterizing the model in terms of $\diffvar^{-1}$ \citep{brownian_diffusion_phylogeography}.
To address the former, \citet{pybus2012} provide an $\order{\ntaxa \ntraits^2}$ dynamic programming algorithm to evaluate (\ref{eq:matrix-normal}) without inversion of the across-taxa variance matrix, similar to \citet{freckleton2012fast}.
This advance certainly makes for more tractable inference under these diffusion models as $\ntaxa$ grows large, but the quadratic dependence on $\ntraits$ still hampers their use for high-dimensional traits. 
Inference can often be slow, taking as long as a day for problems with a dozen traits and about 30 taxa to mix properly \citep{cybis2015assessing}.
Finally, 
direct inference on \(\diffvar\) can often fail to produce a coherent and interpretable conclusion about the number of independent evolutionary processes generating the traits if the matrix cannot be reordered to form approximately separated blocks especially if the signal is too weak to produce many statistically significant cells. 

%
%
%


\subsection{Factor Analysis}
\label{sec:FAmodel}

\newcommand{\normal}[2]{\text{N}\hspace{\minusSpace}\left( #1, #2 \right)}
\newcommand{\gammaDistribution}[2]{\Gamma\hspace{\minusSpace}\left( #1, #2 \right)}

To infer potentially low dimensional evolutionary structure among traits, we rely on dimension reduction via a phylogenetic factor analysis (PFA). 
This model builds on the premise that a small, but unknown number $\nfac \ll \text{min}\left(\ntaxa, \ntraits\right)$ of \textit{a priori} independent univariate Brownian diffusion processes along \(\treeparam\) provides a more  parsimonious description of the covariation in \(\workmat\) than a \(\ntraits\)-dimensional multivariate diffusion. 
We parameterize the PFA in terms of 
an $\ntaxa \times \nfac$ factor matrix $\facmat = (\facmat_1,\ldots,\facmat_{\nfac})$ whose $\nfac$ columns $\facmat_{\idc} = \left(\facmatsm_{1\idc},\ldots,\facmatsm_{\ntaxa \idc}\right)\transpose$ for $\idc = 1,\ldots,\nfac$ represent the unobserved independent realizations of univariate diffusion at each of the $\ntaxa$ tips in $\treeparam$, 
a $\nfac \times \ntraits$ loadings matrix $\loadmat = \{ \loadmatsm_{\idc\idb}\}$ that relates the independent factor columns to $\workmat$, 
and an $\ntaxa \times \ntraits$ model error matrix $\resid$, such that 
\begin{equation}
\begin{aligned}
\workmat &= \facmat \loadmat + \resid .
\end{aligned}
\end{equation}
To inject information about and control for shared evolutionary history $\treeparam$, we specify that 
\newcommand{\arbitraryConstant}{C}
\begin{equation}
\begin{aligned}
\facmat &\sim \mn{
\mathbf{0}}{ 
\facprecTree + \pss^{-1} \mathbf{J}}{
\mathbf{I}_{\nfac}}, \text{ and} \\
\resid &\sim \mn{\mathbf{0}}{\colvarresid}{\rowvarresid^{-1}},
\label{eq:factorModel}
\end{aligned}
\end{equation}
where $\mathbf{I}_{\left( \cdot \right)}$ is the identity matrix of appropriate dimension and the residual column precision $\rowvarresid$ is a diagonal matrix with entries $(\rowvarresidsm_1,\ldots,\rowvarresidsm_\ntraits)$.
Lastly, since $\nfac$ is unknown, we place a reasonably conservative $\text{zero-truncated-Poisson}$  prior on it, such that $\density{\nfac = 1} = 1/2$. 

To better appreciate the details of the PFA model, we briefly compare it to a typical Bayesian factor analysis.
Typical factor analyses assume that all entries of $\facmat$ are independent and identically distributed (iid) as $\normal{0}{1}$, normal random variables with mean $0$ and variance $1$.
In PFA, the shared evolutionary history $\treeparam$ specifies the correlation structure within the $\ntaxa$ entries of column $\facmat_{\idc}$.  
Often, one refers to a given column as a ``factor." 
Across factors, the column variance remains $\mathbf{I}_{\nfac}$ to reflect our assertion that the underlying evolutionary processes generating $\facmat_{\idc}$ are independent of each other. Note that in this model the number of parameters undergoing Brownian Diffusion is assumed to be of dimension \(\nfac\) as opposed to of dimension \(\ntraits\) in the previous model. 

To complete model specification of the loadings $\loadmat$ and residual error $\resid$, we assume
\begin{equation}
\begin{aligned}
\loadmatsm_{\idc\idb} &\sim \normal{0}{1} \text{ for all } \idc \leq \idb, \\
\rowvarresidsm_\idb &\sim \gammaDistribution{\rowvarresidshape}{\rowvarresidrate} \text{ for all trait } \idb \text{ continuous}, \text{ and}
\end{aligned}
\label{eq:priorLoad}
\end{equation}
otherwise $\rowvarresidsm_\idb  = 1$ to preserve identifiability under the scale-free latent model for discrete traits.
Here, 
$\gammaDistribution{\rowvarresidshape}{\rowvarresidrate}$ signifies a gamma distributed random variable with hyperparameter scale $\rowvarresidshape$ and rate $\rowvarresidrate$.

Without further restrictions on $\loadmat$, any factor analysis remains over-specified.
\newcommand{\orthogonalMatrix}{\mathbf{T}}
For example, given an orthogonal $\nfac \times \nfac$ matrix $\orthogonalMatrix$, one may rotate $\facmat$ in one direction and $\loadmat$ in the other and arrive at the same data likelihood, since $\facmat \loadmat = \facmat \orthogonalMatrix \orthogonalMatrix\transpose \loadmat$.
To address this identifiability issue, we fix
lower triangular entries $\loadmatsm_{\idc\idb} = 0$ for $\idc > \idb$ \citep{GewekeZhou1996,Aguilar00bayesiandynamic}.
It is also standard practice to apply the restriction $\loadmatsm_{\idc\idc} > 0$, since otherwise $\facmat \loadmat = (\text{-}\facmat)(\text{-}\loadmat)$. 
While the constraint yields an identifiable posterior distribution with respect to $\facmat$ and $\loadmat$, we do not pursue it here because it introduces bias into our scientific inference on $\loadmat$ and, instead, search for an alternative.

\par

The diagonal and upper triangular entries $\loadmatsm_{\idc\idb}$ for $\idc \le \idb$ of the loadings $\loadmat$ inform the magnitude and effect-direction that the evolutionary process  
captured in factor $\facmat_{\idc}$ contributes to trait $\idb$.
It is possible, and we would argue likely, that $\facmat_{\idc}$ has little or no influence on the trait arbitrarily labeled $\idc$, such that most of the posterior mass of $\loadmatsm_{\idc\idc}$ lies around and close to $0$.
Artificially restricting $\loadmatsm_{\idc\idc} > 0$ forces all of this mass above $0$, signifying a positive association with prior, and hence posterior, probability $1$.

To combat this bias, we recouch these identifiability conditions as a label switching problem in a mixture model and propose a \textit{post hoc} relabeling algorithm \citep{Relabeling3}.
We require $\nfac$ sign constraints, one for each column-row outer-product in forming $\facmat \loadmat$, for posterior identification.
In our prior, we modify Equation (\ref{eq:priorLoad}) to further assign one non-zero entry $\loadmatsm_{\idc\idb} > 0$ per row, but do not specify which one; this assignment mirrors the mixture model labeling.
Hence, we allow the data, not an arbitrary decision, to determine which entry per row reflects a positive association with probability $1$, decreasing potential bias.


\par

Recalling that continuous traits are standardized in $\workmat$ to have mean $0$ and variance $1$ affords several benefits.
First, we can posit a $\mathbf{0}$-matrix mean for $\facmat$ in Equation (\ref{eq:factorModel}) without loss of information.
But, more importantly, when we draw inference on \(\rowvarresid\), we can interpret traits which have precision elements that demonstrate considerable posterior mass at or below 1
to be described insufficiently by the model, since the factors provide no insight beyond a random normal model. 
A third advantage is that standardization helps us select reasonable scales for the non-zero entries in $\loadmat$, namely that these have variance $1$, and hyperparameters for $\rowvarresid$, specifically that $\frac{\rowvarresidshape}{\rowvarresidrate} = 1$.
In practice, \(\rowvarresidshape = \frac{1}{3}\) and \(\rowvarresidrate = \frac{1}{3}\) for analyses in this paper. While these hyperparameter choices are by no means perfect we feel that, under the paradigm of data scaling, they are reasonable and generalizable across a variety of problems.

This model is a simplified form of the item factor analysis models that are described by \cite{IFAPolisci} in the political science literature and \cite{BayesIRT} in the psychology literature with a tree as a prior on the factors instead of an independent normal distribution. In fact, the methods for treating binary and ordinal data described in \cite{IFAPolisci} are the same as those described in \cite{cybis2015assessing}, making for a convenient adaptation of this factor analysis model to phylogenetics using existing software in BEAST.

\subsection{Inference}

\newcommand{\dx}{\mbox{d}}
\newcommand{\indicator}[2]{\ensuremath{\mathbf{1}(#1 \,|\,#2)}}

\label{sec:inference}
Given the trait measurements $\datamat$ and aligned sequences $\sequence$, we strive to learn about the joint posterior distribution of the 
number of evolutionary processes $\nfac$, 
factors $\facmat$, 
loadings $\loadmat$, 
column precisions $\rowvarresid$,
latent trait cut-points $\allCutpoint$
 and evolutionary history $\treeparam$
\begin{equation}
\begin{aligned}
\label{eq:posterior}
\cdensity{\nfac, \facmat, \loadmat, \rowvarresid, \allCutpoint, \treeparam}{\datamat, \sequence}
& \propto 
\cdensity{\datamat}{\nfac, \facmat, \loadmat, \rowvarresid, \allCutpoint} \times
\cdensity{\facmat}{\nfac, \treeparam} \times 
\cdensity{\treeparam}{\sequence} 
\\ 
& \hspace{7em}
\times \cdensity{\loadmat}{\nfac} \times
\density{\rowvarresid} \times
\density{\allCutpoint} \times
\density{\nfac}
\\ 
& = 
\left(
\int
\cdensity{\datamat}{\workmat, \allCutpoint}
\cdensity{\workmat}{\nfac,\facmat, \loadmat, \rowvarresid}
\dx \workmat
\right) 
\cdensity{\facmat}{\nfac, \treeparam} \times 
\cdensity{\treeparam}{\sequence} 
\\ 
& \hspace{7em}
\times \cdensity{\loadmat}{\nfac} \times
\density{\rowvarresid} \times
\density{\allCutpoint} \times
\density{\nfac}
,
\end{aligned}
\end{equation}
where $\cdensity{\datamat}{\workmat, \allCutpoint} \propto \indicator{\datamat}{\workmat, \allCutpoint}$ is the indicator function that the restrictions in Equation (\ref{eq:restriction}) hold.
We accomplish this inference
through 
MCMC,
using a random-scan Metropolis-within-Gibbs scheme \citep{liu1995covariance} for fixed $\nfac$ and a modification of path sampling to then estimate the marginal posterior $\cdensity{\nfac}{\datamat, \sequence}$.
%
For fixed $K$, our Metropolis-within-Gibbs scheme employs transition kernels described in \citet{cybis2015assessing} and references therein to integrate over the evolutionary history \(\treeparam\) and unobserved, latent traits $\workmatsm_{\ida\idb}$ and cut-points $\allCutpoint_{\idb}$ where trait $\idb$ is discrete.  

\par

\newcommand{\deltaIndicator}[1]{
	\mathbf{e}
	_{#1}
}

\newcommand{\gibbsMean}[2]{\mathbf{M}^{
\left( 
#1 
\right)
}_{ #2 }}
\newcommand{\gibbsVariance}[2]{\mathbf{V}^{
\left( 
#1 
\right)
}_{ #2 }}
\newcommand{\idp}{k'}

\newcommand{\gibbsMeanBeta}[2]{\mathbf{M}\hspace{-0.2em}\left( \pathParameter \right)^{
\left( 
#1 
\right)
}_{ #2 }}

\newcommand{\gibbsVarianceBeta}[2]{\mathbf{V}\hspace{-0.2em}\left( \pathParameter \right)^{
\left( 
#1 
\right)
}_{ #2 }}

Here, we focus on transition kernels within the scheme to integrate over the factors \(\facmat\), loadings \(\loadmat\) and residual column precision \(\rowvarresid\).
\citet{LopesWest} derive full conditional distributions for the columns of $\loadmat$ and diagonals of $\rowvarresid$ under a traditional factor analysis.
These full conditional distributions do not change under a PFA and we use them for Gibbs sampling.  
Specifically, for column $\idb$ of $\loadmat$, the first $\idp = \text{min}\left(\idb, \nfac \right)$ entries are non-zero and, given all other random variables, distributed according to a multivariate normal (MVN)
\begin{equation}
\left( \loadmatsm_{1\idb}, \ldots, \loadmatsm_{\idp\idb} \right)\transpose
\, | \,
\workmat, \facmat, \rowvarresid
\sim 
\mvn{
\gibbsMean{\loadmat}{\idb}
}{ 
\gibbsVariance{\loadmat}{\idb}
} \text{ for } j = 1,\ldots,\ntraits
,
\end{equation}
parameterized in terms of its mean 
\begin{align}
\gibbsMean{\loadmat}{\idb} 
&
=
	\gibbsVariance{\loadmat}{\idb} \rowvarresidsm_\idb\facmat_{1:\idp}\transpose\workmat \, \deltaIndicator{\idb} 
\intertext{and variance}
\gibbsVariance{\loadmat}{\idb} 
&
=
	\left(
		\rowvarresidsm_\idb\facmat_{1:\idp}\transpose\facmat_{1:\idp} + \mathbf{I}_{\idp}
	\right)^{-1} 
%
,
\end{align}
where $\facmat_{1:\idp} = \left(\facmat_{1}, \ldots, \facmat_{\idp} \right)$ is the first $\idp$ columns of $\facmat$ and $\deltaIndicator{\idb}$ is the unit-vector in the direction of trait $\idb$.
Further,
\begin{equation}
\rowvarresidsm_{\idb} 
\, | \, \workmat, \facmat, \loadmat
\sim 
\gammaDistribution{
	\rowvarresidshape + \frac{\ntaxa}{2}
}{
\rowvarresidrate + \frac{1}{2} \,
	\deltaIndicator{\idb}\transpose
	\left( 			
		\workmat - \facmat \loadmat
	\right)\transpose
	\left(
		\workmat - \facmat \loadmat
	\right)
	\deltaIndicator{\idb}
},
\end{equation}
if trait $\idb$ is continuous.
Appendix \ref{app:gibbs} provides
derivations of these full conditional distributions.
%
%
Gibbs sampling all columns of $\loadmat$ carries a computation order $\order{\ntaxa \nfac^2 \ntraits}$, arising from the matrix multiplication of $\facmat_{1:\idp}\transpose\facmat_{1:\idp}$ for each trait.  
The matrix inversion is not rate-limiting here since $\ntaxa \gg \nfac$.
Likewise, Gibbs sampling $\rowvarresid$ remains very light-weight at $\order{\ntaxa \nfac \ntraits}$, stemming from the sparse multiplication of $\facmat\loadmat\deltaIndicator{\idb}$ for each trait.
While we write that the order of both Gibbs samplers depend on $\ntraits$ to be clear that we must iterate over all traits, the astute reader has already recognized the conditional independence of updates between traits, such that we may execute updates for each trait in parallel. 


The traditional Gibbs sampler for $\facmat$ fails in the phylogenetic setting for more than a handful of taxa, since determining the full conditional distribution of $\facmat$ requires inverting the matrix $\left(\facprecTree + \pss^{-1} \mathbf{J}\right)$.
As mentioned previously, but worth repeating, this task stands as prohibitive with a computational order $\order{\ntaxa^3}$ and presents a major challenge for PFA.

We circumvent this difficulty by exploiting the structure of the phylogenetic tree $\treeparam$.
Probability models on directed, acyclic graphs lend themselves well to dynamic programming for determining marginalized data likelihoods, such as Felsenstein's pruning algorithm for sequence data \citep{felsenstein1973} and related work for Brownian diffusion \citep{pybus2012}, and conditional predictive distributions, like those obtained for (ancestral) sequence reconstruction.
\newcommand{\rowSet}{_{\ida \cdot}}
\newcommand{\notRowSet}{_{\text{-}\ida \cdot}}
\def\facmeancond{\boldsymbol{\mu_{\facmat\notRowSet}}}
\def\facpreccond{\boldsymbol{\Lambda_{\facmat\notRowSet}}}

In extending these conditional distributions to Brownian diffusion, first let 
$\facmat\rowSet = ( \facmatsm_{\ida 1}, \ldots, \facmatsm_{\ida \nfac} )$ identify row $\ida$ of $\facmat$, more specifically all latent factor values attributed to taxon $\ida$, and let $\facmat\notRowSet$ concatenate the remaining rows.
Given that $\facmat$ is matrix-normally distributed with an across-taxa (row) variance that depends on the phylogeny $\treeparam$, \citet{cybis2015assessing} provide a tree-traversal-based algorithm to determine 
$\cdensity{\facmat\rowSet}{\facmat\notRowSet, \treeparam}$ that remains a
multivariate normal distribution.
The algorithm requires first a post-order tree-traversal to determine the joint distribution of all tip-values descendent to each internal node and then a pre-order tree-traversal back to taxon $\ida$ to compute its prior conditional mean $\facmeancond$ and precision $\facpreccond$.
Since the across-factor (column) variance on $\facmat$ is diagonal, the dynamic programming algorithm runs quickly in $\order{\ntaxa \nfac}$. 
Using this result, we determine the full conditional distribution
\begin{equation}
\facmat\rowSet\transpose \,| \,
	\workmat,
	\facmat\notRowSet,
	 \loadmat, 
	 \rowvarresid,  
	 \treeparam
		\sim 
\mvn{
\gibbsMean{\facmat}{\ida}
}{ 
\gibbsVariance{\facmat}{\ida}
}
\text{ for } \ida = 1,\ldots,\ntaxa,
\label{eq:factFull}
\end{equation}
with mean
\begin{align}
\gibbsMean{\facmat}{\ida} &= \gibbsVariance{\facmat}{\ida}
	\left(
		\loadmat
		\rowvarresid
		\workmat\transpose
		\deltaIndicator{\ida}
		+ 
		\facpreccond\facmeancond 
	\right) \label{eq:factMean} \\
\intertext{and variance}
\gibbsVariance{\facmat}{\ida} &= 				
	\left(
		\loadmat\rowvarresid\loadmat\transpose + \facpreccond
	\right)^{-1} , \label{eq:factVar}
\end{align}
where $\deltaIndicator{\ida}$ is the unit-vector in the direction of taxon $\ida$.
Appendix 
\ref{app:gibbs}
delivers a derivation of this full conditional distribution. The evaluation of this full conditional distribution runs in $\order{\nfac^2\ntraits},$ where the term $\loadmat \rowvarresid \loadmat \transpose$ is rate limiting.

Employing Equations (\ref{eq:factFull}) - (\ref{eq:factVar}), we can cycle over $\ida$ to fabricate a tractable Gibbs sampler for $\facmat$ with total computational order $\order{\ntaxa^2 \nfac + \ntaxa\nfac^2\ntraits}$.
It is fruitful to compare this work with the rate-limiting step for inference under the non-sparse model.
Here, sampling the precision matrix $\diffvar^{-1}$ carries a computational cost of $\order{\ntaxa \ntraits^2}$.
From these bounds, it is clear that increasing numbers of taxa $\ntaxa$ should limit PFA, while increasing numbers of traits $\ntraits$ should limit the non-sparse model from a computational work per MCMC iteration perspective.
However, per-iterative arguments ignore the posterior correlation between model parameters and its influence on MCMC mixing times. 

\newcommand{\idMax}[1]{\idb_{#1}}
\newcommand{\itnum}{m}
\newcommand{\iteration}{^{(\itnum)}}
\newcommand{\itplusone}{^{(\itnum+1)}}

Finally, to maintain identifiability with respect to $\facmat$ and $\loadmat$ in the posterior, we propose a simple \textit{post hoc} relabeling algorithm \citep{Relabeling3}.
We sample $(\facmat\iteration, \loadmat\iteration)$ from  $\cdensity{\nfac, \facmat, \loadmat, \rowvarresid, \allCutpoint, \treeparam}{\datamat, \sequence}$ for MCMC iteration $\itnum = 1,\ldots,M$ assuming a sign-unconstrained prior.
From this unconstrained sample, we select for each row $\idc$ in $\loadmat$ 
the column element with the fewest number of sign changes between iterations.
Assume for row $\idc$, this is column $\idMax{\idc}$. 
We then constrain our sample by multiplying $\facmat_{\idc}\iteration$ 
and row $\idc$ of $\loadmat\iteration$ by the sign of $\loadmatsm_{\idc \idMax{\idc}}\iteration$.
No further sample reweighing is necessary because 
$\cdensity{\facmat}{\nfac, \treeparam} = \cdensity{\text{-}\facmat}{\nfac, \treeparam}$ is also invariant to reflection.

\paragraph{Model Selection}

\newcommand{\smallFactorCount}{S}

To estimate the marginal posterior density $\cdensity{\nfac}{\datamat, \sequence}$, we rely on a variant of path sampling that we equip to successfully integrate latent variable $\workmat$ when traits are discrete.  
We employ our variant to approximate each marginal likelihood $\cdensity{\datamat, \sequence}{\nfac = \idc}$ for $\idc = 1,\ldots,\smallFactorCount$, where $\smallFactorCount$ is a relatively small number such as $\min\{ \ntraits, 10 \}$, after which we approximate $\cdensity{\datamat, \sequence}{\nfac > \smallFactorCount} = 0$.
Then, invoking Bayes theorem, $\cdensity{\nfac = \idc}{\datamat, \sequence} \propto \cdensity{\datamat, \sequence}{\nfac = \idc} \density{\nfac = \idc}$.
Moreover, through this approach, we can address the model selection problem of how many independent factors do the data support through Bayes factors \citep{Jeffreys35}:
\begin{equation}
\frac{
	\cdensity{\nfac = \idc}{\datamat, \sequence}
}{
	\cdensity{\nfac = \idc'}{\datamat, \sequence}
} =
\frac{
	\cdensity{\datamat, \sequence}{\nfac = \idc}
}{
	\cdensity{\datamat, \sequence}{\nfac = \idc'}
}
\frac{
	\density{\nfac = \idc}
}{
	\density{\nfac = \idc'}
} .
\end{equation}
\citet{LopesWest} and \citet{Ghosh09defaultprior} have been strong proponents of Bayes factors to determine the optimal number of factors in a traditional factor analysis, where \citet{LopesWest} employ a simple harmonic mean estimator \citep{HMEsampling} to estimate their marginal likelihoods.
This estimator performs poorly in highly structured phylogenetic models and path sampling has largely supplanted it \citep{BeastPathSampling1}.

\newcommand{\jlikelihood}[1]{\ensuremath{l(#1 )
}}
\newcommand{\clikelihood}[2]{\ensuremath{l(#1 \,|\,#2)}}
\newcommand{\working}[1]{\ensuremath{\hat{p}(#1 )
}}
\newcommand{\cworking}[2]{\ensuremath{\hat{p}(#1 \,|\,#2)}}
\newcommand{\wtf}[1]{\ensuremath{h(#1 )
}}
\newcommand{\cwtf}[2]{\ensuremath{h(#1 \,|\,#2)}}
\newcommand{\MASpath}[1]{\ensuremath{q(#1)}}

Path sampling is an MCMC-based integration technique to estimate marginal likelihoods, such as $\cdensity{\datamat, \sequence}{\nfac}$.
The technique constructs a series of power posteriors \citep{friel2008marginal} at various temperatures $\pathParameter \in [0,1]$, where $\beta = 1$ corresponds to a joint density $\clikelihood{\datamat, \sequence, \workmat, \facmat, \loadmat, \rowvarresid, \allCutpoint}{\nfac}$ proportional, but with an unknown constant, to $\cdensity{\datamat, \sequence}{\nfac}$ and $\beta = 0$ yields a normalized density $\cworking{\workmat, \facmat, \loadmat, \rowvarresid, \treeparam, \allCutpoint}{\nfac}$ that does not depend on the data, often a combination of the prior and other working distributions, see e.g.~\citep{baele2016genealogical}. 
The usual power posterior path is $\MASpath{\pathParameter, \datamat, \sequence, \paramList} = \jlikelihood{\datamat, \sequence, \paramList}^{\pathParameter} \times \working{\paramList}^{1 - \pathParameter}$, where $\paramList$ is the set of all parameters in the model we are considering. For example, in PFA, $\paramList = \left\{\workmat, \facmat, \loadmat, \rowvarresid, \treeparam, \allCutpoint\right\}.$

\newcommand\integratedPrior[1]{\hat{P}\left(#1\right)}
\newcommand\integratedPath[1]{Q\left(#1\right)}
\newcommand{\domain}{\boldsymbol{\Omega}}

In latent models with discrete traits, however, the support of the latent variable $\workmat$ changes when the data are observed \citep{heaps2014computation}.
In particular, our unnormalized joint density $\jlikelihood{\datamat, \sequence, \paramList}$ is zero for values of $\workmat$ that are incompatible with $\datamat$ because $\cdensity{\datamat}{\workmat, \allCutpoint} = 0$,
therefore a trait $\workmatsm_{\ida\idb}$ only has support over $(\cutpoint_{\ida(c-1)}, \cutpoint_{\ida c}]$ if $\datamatsm_{\ida\idb} = c$, 
while $\working{\cdot}$ places non-zero density over all possible values $\workmatsm_{\ida\idb} \in (-\infty, \infty)$. Our working distribution, for example, assumes $\workmatsm_{\ida\idb} \sim N(0, 1)$ when $\workmatsm_{\ida\idb}$ is random.
If we factor $\jlikelihood{\datamat, \sequence, \paramList}$ into a support condition $\indicator{\datamat}{\workmat,\allCutpoint}$
 and the remaining likelihood $\wtf{\datamat, \sequence, \paramList}$, then the standard path used in this scenario \citep{heaps2014computation} is 
\begin{equation}
\MASpath{\pathParameter, \datamat, \sequence, \paramList} = 
 \indicator{\datamat}{\workmat,\allCutpoint}
\times
\wtf{\datamat, \sequence, \paramList}^{\pathParameter} 
\times
\working{\paramList}^{1 - \pathParameter} .
\label{eqn:heapspath}
\end{equation}
For the power posterior method to yield the marginal likelihood $\cdensity{\datamat}{\nfac},$ it is necessary \citep{friel2008marginal} that
\begin{equation}
\label{eqn:integral}
\int \left\{\lim_{\pathParameter\to0}\MASpath{\pathParameter, \datamat, \sequence, \paramList} \right\}\dx\paramList = 1 .
\end{equation}
Plugging (\ref{eqn:heapspath}) into (\ref{eqn:integral}), we find 
\begin{equation}
	\int \left\{\lim_{\pathParameter\to0}\MASpath{\pathParameter, \datamat, \sequence, \paramList} \right\} \dx\paramList = \int
 \indicator{\datamat}{\workmat,\allCutpoint}
\times
\working{\paramList} \dx \paramList.
\label{eqn:limit}
\end{equation}
If we define $\domain$ as the region where $\indicator{\datamat}{\workmat,\allCutpoint} = 1,$ then we see that
\begin{equation}
\int_{\domain} 
\working{\paramList} \dx \paramList
< 1,
\end{equation}
since $\domain \subsetneq$ the support of $\paramList.$
While it is theoretically possible to construct $\working{\paramList}$ such that it is normalized to 1 over $\domain$, previous attempts to do so have failed. Alternatively, \cite{heaps2014computation} attempt to approximate such a distribution by fixing $\allCutpoint$ and ignoring the corresponding integral.

\newcommand\modPath[1]{q^*\left(#1\right)}
\newcommand\modIntPath[1]{Q^*\left(#1\right)}

We posit an exact solution by proposing a new path that relies on a softening threshold.
Consider the modified path
\begin{equation}
\modPath{\pathParameter, \datamat, \sequence, \paramList} = 
\left\{ 
1 - 
	\left[
	1 - \indicator{\datamat}{\workmat,\allCutpoint}
	\right]
\pathParameter
\right\}
\times
\wtf{\datamat, \sequence, \paramList}^{\pathParameter} 
\times
\working{\paramList}^{1 - \pathParameter} .
\label{eq:newPath}
\end{equation}
Following from (\ref{eqn:integral}), we find that
\begin{equation}
\int
\left\{ 
\lim_{\pathParameter\to0}\modPath{\pathParameter, \datamat, \sequence, \paramList} 
\right\}
\dx\paramList = 
\int \working{\paramList} \dx\paramList = 1,
\end{equation}
by construction.

Lastly, in order to adapt the power posterior method, at each step in the series we need to compute the derivative of 
$\log \modPath{\pathParameter, \datamat, \sequence, \paramList}$ with respect to $\pathParameter$.
From Equation (\ref{eq:newPath}), we see that
\begin{align}
\frac{\partial}{\partial \pathParameter}
\log \modPath{\pathParameter, \datamat, \sequence, \paramList} = - \frac{
	1 - \indicator{\datamat}{\workmat,\allCutpoint}
}{
1 - \left[ 
1 - \indicator{\datamat}{\workmat,\allCutpoint}
\right] \pathParameter
}
 + \log \wtf{\datamat, \sequence, \paramList} - \log \working{\paramList},
\end{align}
and observe that there is no singularity at $\pathParameter = 1$ since, at that point in the path, latent variable $\workmat$ only assumes values in $\domain$, such that $\indicator{\datamat}{\workmat,\allCutpoint} = 1$.

\section{Empirical Examples}

\subsection{Columbine Flower Development}

Columbine genus \textit{Aquilegia} flowers have attracted at least three different pollinators across their evolutionary history: bumblebees (Bb), hawkmoths (Hm) and hummingbirds (Hb).
\citet{whittall2007} question the role that these pollinators play in the tempo of columbine flower evolution, tracked through the color, length and orientation of different anatomical floral features, and are particularly interested in how transitions between pollinators relate to spur length.
\citet{cybis2015assessing} take up this question by examining $\ntraits = 12$ different traits for $\ntaxa = 30$ monophyletic populations from the genus \textit{Aquilegia} that include $10$ continuously valued traits, a binary trait that indicates presence or absence of anthocyanin pigment and a final ordinal trait indicating the primary pollinator for that population. 
\citet{whittall2007} propose a Bb-Hm-Hb ordering and we use the fixed phylogenetic tree the authors employ in their analysis.
Through fitting a latent multivariate Brownian diffusion (LMBD) model parameterized in terms of a $12 \times 12$ variance matrix $\diffvar$, \citet{cybis2015assessing} find the data strongly support the proposed ordering over alternative orderings.
We return to the relationship between pollinator and the other traits and test whether a PFA returns a better understanding of the evolutionary factors driving their interrelated change compared to an LMBD model.

\begin{table}
\caption{Log marginal likelihood estimates for the number $\nfac$ of independent factors driving evolution under a phylogenetic factor analysis (PFA) and a latent multivariate Brownian diffusion (LMBD) model in \textit{Aquilegia}, and \textit{Poeciliidae} and multivariate Brownian diffusion (MBD) in \textit{Balistidae}. 
%
The $\nfac=2$ model for \textit{Aquilegia}, the $\nfac=3$ and $\nfac=4$ model for \textit{Poeciliidae} and the $\nfac=5$ model for \textit{Balistidae} achieve the highest marginal likelihoods. 
\label{tab:bf}}
\begin{center}
\begin{tabular}{ccD{.}{.}{-1}D{.}{.}{-1}}
& & \multicolumn{1}{c}{Log marginal} \\
& \multicolumn{1}{l}{Model} & \multicolumn{1}{c}{likelihood}
\\ 
\hline
\\
 \parbox[t]{2mm}{\multirow{4}{*}{\rotatebox[origin=c]{90}{\textit{Aquilegia}}}}
& $\nfac = 1$ & -385.4 \\
& $\nfac = 2$ & -366.9  \\
& $\nfac = 3$ & -374.3 \\
& LMBD         & -391.1 & \\ 
\\
\hline
\\
& $\nfac = 2$ & -536.0  \\ 
\parbox[t]{2mm}{\multirow{4}{*}{\rotatebox[origin=c]{90}{ \textit{Poeciliidae}}}}
& $\nfac = 3$ & -500.7  \\ 
& $\nfac = 4$ & -501.0 \\
& $\nfac = 5$ & -505.9  \\ 
& LMBD        & -592.3 & \\ 
\\
\hline
\\
 \parbox[t]{2mm}{\multirow{4}{*}{\rotatebox[origin=c]{90}{ \textit{Balistidae}}}}
& $\nfac = 4$ & -15622.0  \\
& $\nfac = 5$ & -15603.5  \\ 
& $\nfac = 6$ &  -15610.4 \\
& MBD        & -15673.2 \\
\\
\hline
\end{tabular}
\end{center}
\end{table}

Under our PFA, the most probable number of independent evolutionary processes is $\nfac = 2$, with a log Bayes factor $>7$ over the neighboring $K = 1$ or $K = 3$ factor parameterizations (Table \ref{tab:bf}).
Further, the PFA with $\nfac=2$ is favored over the LMBD model with a log Bayes factor $>24$ when assuming equal prior probabilities over these two models.

\begin{figure}
\centering
    \includegraphics[width=\textwidth]
    {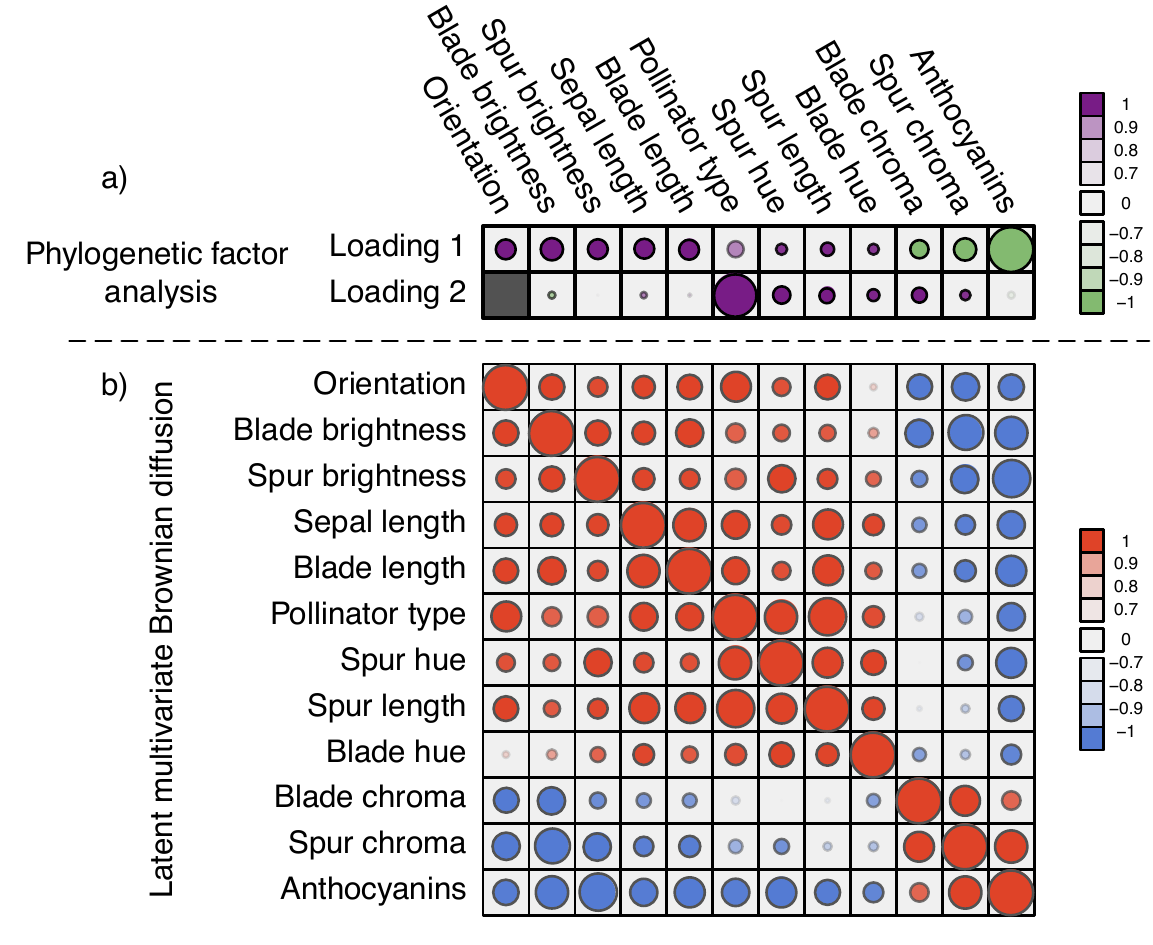}
	\caption{Processes driving columbine flower evolution inferred through phylogenetic factor analysis (PFA) or latent multivariate Brownian diffusion (LMBD).
	 a) Loadings $\loadmat$ estimates from a $\nfac = 2$ factor PFA model. 
Purple circles represent traits positively associated with traits represented by other purple circles within a loading, and negatively associated with traits represented by green circles within a loading. Similarly, traits represented by green circles are positively associated with traits represented by green circles within a loading. Size represents the magnitude of the value of the loadings. Opacity represents the posterior probability that the sign of the given element is equal to the sign of the posterior mean. The greyed out cell represents a structural 0 introduced for identifiability reasons.
%
	The magnitude for anthocyanins and pollinator type is less relevant since those measurements are discrete. 
	b) Correlation matrix estimate from a LMBD model. 
	Red represents positive correlation, blue represents anti-correlation, and opacity represents the probability that the sign of the element is equal to the sign of the posterior mean. 
	Size of the circle represents the magnitude of the correlation.
The PFA captures well two independent processes, while the LMBD groups these processes together.	
	}
	\label{fig:columbineResults}
\end{figure}

The PFA has high explanatory power for all continuous traits (Table \ref{tb:prec}) and
Figure \ref{fig:columbineResults} presents our inference on the relationships between traits under the  PFA with $\nfac = 2$ and compares these findings to inference under the LMBD model.
The first evolutionary process $\facmat_{1}$ approximately partitions the traits into two groups.
One group includes: orientation, blade brightness, spur brightness, sepal length, blade length, pollinator type, spur hue, spur length, blade hue, and expected trait values increase (displayed loadings entries $\loadmatsm_{\idc\idb}$ in purple) as the factor grows over the phylogeny.
The other group includes: blade chroma, anthocyanins pigment presence and, with less posterior probability, spur chroma, and expected trait values decrease (green) as the factor grows.
A possible exception to the $\facmat_{1}$ partitioning is the pollinator trait, where we estimate only a 92\% posterior probability that this cell has the same sign as its posterior mean.

Ignoring the uncertainty in pollinator trait inclusion for the moment, this partitioning recapitulates the block structure that \citet{cybis2015assessing} report using an LMBD model and an arbitrary thresholding on the posterior mean estimates of the individual pairwise correlation entries in $\diffvar$.
However, in Figure \ref{fig:columbineResults} we quantify the LMBD uncertainty by shading our inference using the same probability measure as we do for our PFA model.
Taking correlation uncertainty into consideration we see that, for example the LMBD model would assert that there is no correlation between blade chroma and spur hue. 
The PFA model by contrast offers the more nuanced assessment that these traits are related through two independent underlying processes, one process of which has a positive association between these traits, the other of which has a negative association. 

\begin{table}
\caption{Precision $\boldsymbol{\Lambda}$ posterior mean and 95\% Bayesian credible interval estimates under the latent factor model for the traits in \textit{Aquilegia}, in \textit{Poeciliidae} and in \textit{Balistidae}. The PFA model explains all of the continuous traits in these models better than a $N(0,1)$ distribution on the standardized traits.}
\begin{center}
\begin{tabular}{crD{.}{.}{-1}D{.}{.}{-2}ccccccc}
 & & \multicolumn{1}{c}{Posterior}  &
\multicolumn{1}{c}{\hspace{.4in}95\% Bayesian }
\\ & \multicolumn{1}{c}{Trait} & \multicolumn{1}{c}{mean}&  \multicolumn{1}{c}{\hspace{.4in}credible interval} \\
\hline
\\
& Orientation & 2.1 & [1.0, ~3.3]  \\
& Spur length & 4.4 & [2.0, ~7.1] \\
& Blade length & 3.0 & [1.4, ~4.8]  \\
\parbox[t]{2mm}{\multirow{4}{*}{\rotatebox[origin=c]{90}{\textit{Aquilegia}}}}
& Sepal length & 2.6 & [1.3, ~4.1]  \\
& Spur chroma & 4.2 & [1.8, ~6.9]  \\
& Spur hue & 6.2 & [2.6, 10.5] \\
& Spur brightness & 2.7 & [1.2, ~4.3] \\
& Blade chroma & 2.3 & [1.1, ~3.7]  \\
& Blade hue & 2.1 & [1.0, ~3.2] \\
& Blade brightness & 3.3 & [1.4, ~.6] \\ 
\\
\hline
\\
& Matrotrophy index & 14.3 & [5.6, 23.2]  \\
 \parbox[t]{2mm}{\multirow{4}{*}{\rotatebox[origin=c]{90}{ \textit{Poeciliidae}}}}
 \parbox[t]{2mm}{\multirow{4}{*}{\rotatebox[origin=c]{90}{ ($\nfac = 3$)}}}
& Gonopodium length        & 9.3 & [4.3, 16.1]  \\
& Male body length        & 3.5 & [2.4, ~4.6] \\
& Male body weight        & 2.8 & [1.9, ~3.7]\\
& Female body length        & 10.5 & [5.7, 15.5]  \\
& Female body weight        & 15.1 & [8.0, 24.3]  \\ 
\\
\hline
\\
& Matrotrophy index & 13.8 & [5.5, 22.7]  \\
 \parbox[t]{2mm}{\multirow{4}{*}{\rotatebox[origin=c]{90}{ \textit{Poeciliidae}}}}
 \parbox[t]{2mm}{\multirow{4}{*}{\rotatebox[origin=c]{90}{ ($\nfac = 4$)}}}
& Gonopodium length        & 9.1 & [4.4, 15.5]  \\
& Male body length        & 3.5 & [2.3, ~4.8] \\
& Male body weight        & 2.8 & [1.9, ~3.8]\\
& Female body length        & 10.5 & [5.8, 15.5]  \\
& Female body weight        & 14.7  & [8.2, 22.5]  \\
\\
\hline
\end{tabular}
\end{center}
\label{tb:prec}
\end{table}

In addition to improved uncertainty quantification in the block structure of traits, our PFA returns a second independent evolutionary process $\facmat_2$ that relates pollinator with spur length and, in addition, spur and blade chroma and hue, with posterior probability approaching $1$.
The existence of two distinct processes, one of which directly connects pollinator and spur length, sheds additional insight into the original hypothesis that \citet{whittall2007} pose.
The LMBD model fails to pick up on this, in addition to returning a worse fit to the data.
 \newcommand\myspace{\hphantom{1}}

\subsection{Transitions to Placental Reproduction}
\label{sec:Poeciliidae}

The freshwater fish \textit{Poeciliidae} represent a family of model organisms in which one can study the transition from non-placental to placental reproduction and the evolutionary pressures associated with placental introduction.
\cite{fish_evolution} define a matrotrophy index to be the log-ratio of the dry weight of newborn fish to the dry weight of eggs at fertilization as a proxy measure of how reliant a fish species is on its placenta for reproduction.
Using phylogenetic generalized least squares (PGLS) \citep{phylogenetic_least_squares}, \citet{fish_evolution} find that \textit{Poeciliidae}
dichromatism, courtship behavior, superfetation, and a sexual selection index are all correlated over evolutionary history with the matrotrophy index.
Unlike PFA, PGLS as used by \cite{fish_evolution} does not adjust for potential evolutionary relationships between the traits.  
Failure to do so can lead to false positive measures of association between individual traits and the matrotrophy index.

%
%

\citet{fish_evolution} collect from the literature or measure 14 life-history traits and compile from GenBank or sequence 28 different genes across \textit{Poeciliidae} species. 
In our analysis, we only use $\ntraits = 11$ traits since three of the original traits are functions of the included ones.
Of these traits, five are discrete-valued: 
dimorphic coloration (dichromatism), 
courtship behavior, 
superfetation, 
the presence or absence of ornamental display traits and 
a count composite of the presence or absence of three other male behaviors (sexual selection index).
Six are continuous-valued:
log weight and log length for males and females, 
gonopodium length and 
matrotrophy index.
Considering species with at least one trait measurement, there are $\ntaxa = 98$ 
taxa, for which we assume the same fixed phylogenetic tree that \citet{fish_evolution} estimate and similarly condition on in their PGLS analysis.
Importantly, 182 trait measurements remain missing.
We treat these measurements as missing-at-random in our PFA and do not need to further prune the tree or impute values that may further introduce bias.

\begin{figure}
\centering
    \includegraphics[width = \textwidth]
    {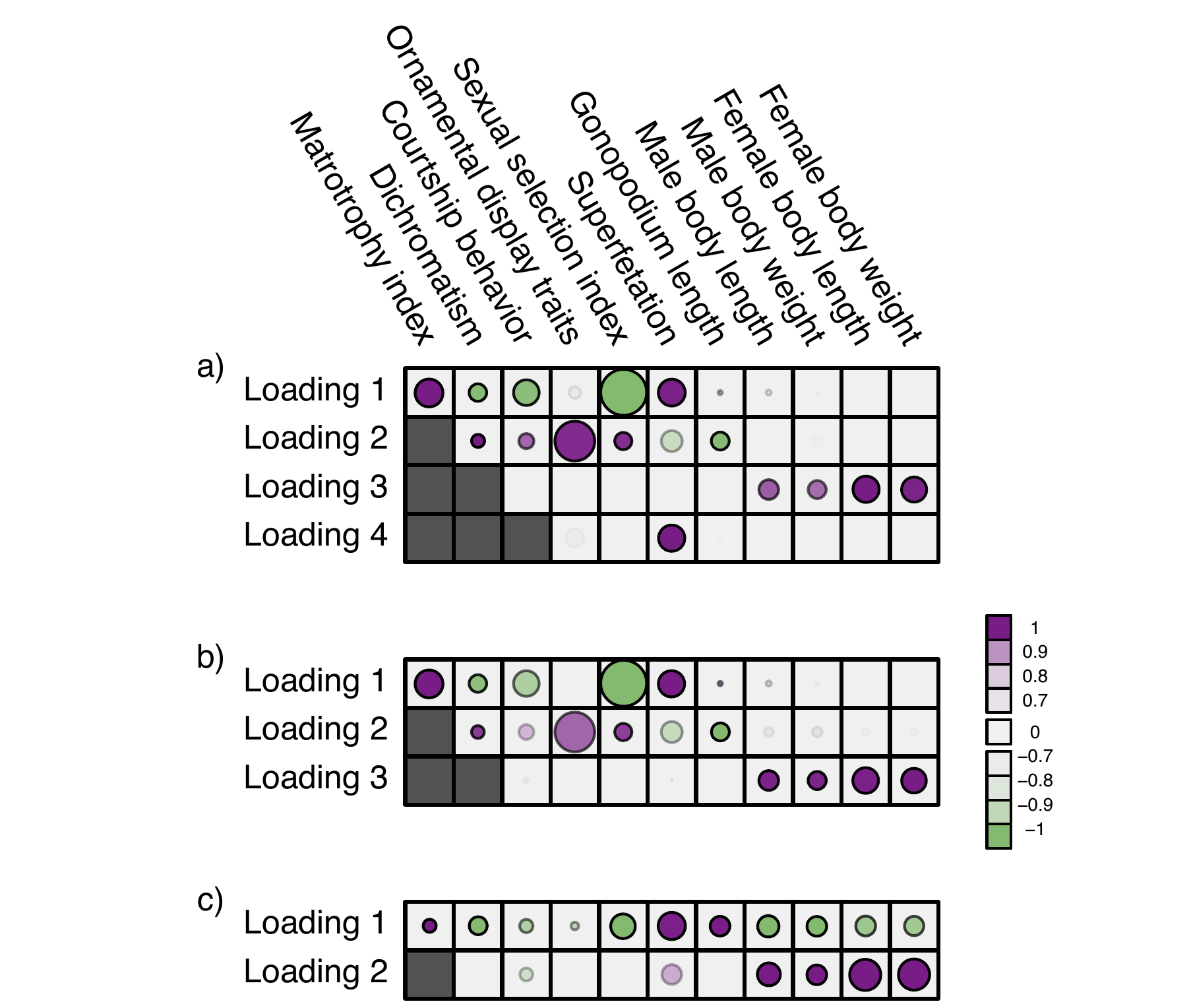}
  \caption{
  Processes driving transitions to placental reproduction inferred through PFAs. Loading $\loadmat$ estimates from the a)  $\nfac = 4$, b) $\nfac = 3$ and c) $\nfac = 2$ factor models.
Loadings size, coloring and density follow those of Figure \ref{fig:columbineResults}.
Note that the magnitude for dichromatism, courtship behavior, ornamental display traits, sexual selection index and superfetation is less relevant since those data are discrete. We include the two factor model for direct comparison to the results of \cite{fish_evolution}.
Loadings in the more probable $\nfac = 3$ and $\nfac = 4$ factor models do not support an association between matrotrophy index and gonopodium length nor body weights and lengths.
%
 %
 }
\label{fig:poeciliidae_results}
\end{figure}

\cite{fish_evolution} find that dichromatism, courtship behavior, superfetation, and sexual selection index are all correlated with the matrotrophy index. 
Figure \ref{fig:poeciliidae_results} shows that this concurs with the results of a $\nfac = 2$ factor PFA.
This small model fit also highlights a weakness of traditional factor analysis assumptions that fix the diagonal elements of the loadings matrix to be positive.
In particular, dichromatism is unrelated to the other traits in the second factor, while the positivity constraint would have forced its inclusion.
However, the most probable number of independent evolutionary processes is $\nfac = 3$ or $\nfac = 4$, with a log Bayes factor in favor $\nfac = 3$ over $\nfac = 2$ of $35.3$
and a log Bayes factor in favor of $\nfac = 4$ over $\nfac = 5$ of $4.9$ (Table \ref{tab:bf}).
Since a log Bayes factor of only $0.3$ separates the $\nfac = 3$ and $\nfac = 4$ models, we include both models in our results, and the data strongly support these PFA models over the LMBD model (log Bayes factor $\approx$ 92).

 \newcommand\evproc[2]{\facmat^{\left(#1\right)}_#2}


Loadings for the independent evolutionary process factors $\evproc{3}{\idc}$ and $\evproc{4}{\idc}$ under the $\nfac = 3$ and $\nfac = 4$ PFA models, respectively, recapitulate a negative assocation between the matrotrophy index and dichromatism, courtship behavior, and sexual selection index, and a positive association with superfetation (Figure \ref{fig:poeciliidae_results}, first loading).  
However, unlike in \cite{fish_evolution}, the PFA does not recover with high posterior probability a relationship between matrotrophy index and gonopodium length nor with body weights and lengths, suggesting that these were false positive findings.
For both PFA models,
second independent processes $\evproc{3}{2}$ and $\evproc{4}{2}$ drive dichromatism, courtship behavior, ornamental display traits and sexual selection index positively and superfetation and gonopodium length negatively.
Both models also identify similar third independent processes $\evproc{3}{3}$ and $\evproc{4}{3}$ relating body lengths and weights.
It is perhaps surprising that these size measurements are unrelated to any of the other reproductive characteristics.
The only marked difference between the $\nfac = 3$ and $\nfac = 4$ factor models exists in the presence of a fourth evolutionary process $\evproc{4}{4}$ in the $\nfac = 4$ factor model that controls the presence or absence of superfetation independently of all other traits.

The precision elements $\boldsymbol{\Lambda}$ for both the $\nfac=3$ and $\nfac=4$ factor models are all significantly greater than 1 and therefore indicate that, for both models, our PFA provides good insight into the relationship of the continuous traits (Table \ref{tb:prec}). 
Further, the precision elements are in broad agreement between the $\nfac=3$ and $\nfac=4$ factor models, as we expect due to the negligible difference in marginal likelihoods.

Frequentist-based factor analysis is only identifiable if the number of parameters inferred for a variance/covariance matrix is greater than the number of parameters that need to be inferred for the factor analysis. 
Interestingly, our PFA model produces interpretable results in spite of the fact that the correlation model has 66 free parameters as opposed to 333 free parameters for the $\nfac = 3$ factor model, and 436 free parameters for the $\nfac = 4$ factor model. 


\subsection{Triggerfish Fin Shape}
\label{sec:triggerfish}

The fish family \textit{Ballistidae}, commonly know as triggerfish, live mostly in reefs; however, the particular part of the reef in which they live can vary.
This variability affects not only their diet, but also their mobility needs that fin shapes well reflect \citep{triggerfish1}.
To model shape changes through evolution, phylogenetic morphometrics often relies heavily on principle components analysis (PCA) \citep{revell2009size,polly2013phylogenetic}.
However, deterministic data reduction via PCA can introduce bias \citep{uyeda2015comparative} and, more importantly, inference of principal components while simultaneously adjusting for an uncertain evolutionary history remains a continuing challenge.
PFA offers an alternative approach.


For $\ntaxa = 24$ triggerfish species, 
\cite{triggerfish1} sequence and align 12S (833 nucleotides, nt) and 16S (563 nt) mitochondrial genes and RAG1 (1471 nt), rhodopsin (564 nt) and Tmo4C4 (575 nt) nuclear genes, and
\cite{triggerfish2} digitally photograph and mark 13 semi-landmark Cartesian coordinates for pectoral, dorsal and anal fins, generating $\ntraits = 78$ measurements per species.
%
Among these morphometric measurements,
the species \textit{Balistapus undulatus} is missing dorsal and anal fins landmarks, and the species \textit{Rhinecanthus assasi} lacks pectoral fin landmarks.
For these, we assume the missing data are missing at random.

\par 

\newcommand{\invariantProp}{p_{\text{\tiny inv}}}
\newcommand{\yangShape}{\alpha}
\newcommand{\stationaryDistribution}{\boldsymbol{\pi}}

To accommodate phylogenetic uncertainty within $\cdensity{\treeparam}{\sequence}$, 
we concatenate gene alignments into $\sequence$ and model nucleotide sequence substitution along the unknown evolutionary history $\treeparam$ through the \cite{Hasegawa85} continuous-time Markov chain with unknown transition:transversion rate ratio $\kappa$ and stationary distribution $\stationaryDistribution
$.
We incorporate across-site rate variation using a discretized, one-parameter Gamma distribution \citep{gammaSiteRate} with unknown shape $\yangShape$ and proportion $\invariantProp$ of invariant sites.
To specify prior $\density{\treeparam, \kappa, \stationaryDistribution, \yangShape, \invariantProp}$, we make relatively uninformative choices, documented in the BEAST extensible markup language (XML) file in the Supplementary Material. 
%

These triggerfish sequences and traits favor the $\nfac = 5$ factor model with a log Bayes factor of $18.5$ over the $\nfac = 4$ factor model and $6.9$ over the $\nfac = 6$ factor model (Table \ref{tab:bf}).
Further, these data favor the $\nfac = 5$ factor model over the multivariate Brownian diffusion (MBD) model with a log Bayes factor of $69.7$.
Even if this support were equivocal, we caution against using a MBD to model these traits.
The unknown variance matrix $\Sigma$ carries $\ntraits (\ntraits + 1) / 2 = 3081$ degrees-of-freedom that dwarfs the $\ntaxa \times \ntraits = 1872$ possible measurements.


\begin{figure}
\centering
    \includegraphics[width = \textwidth]{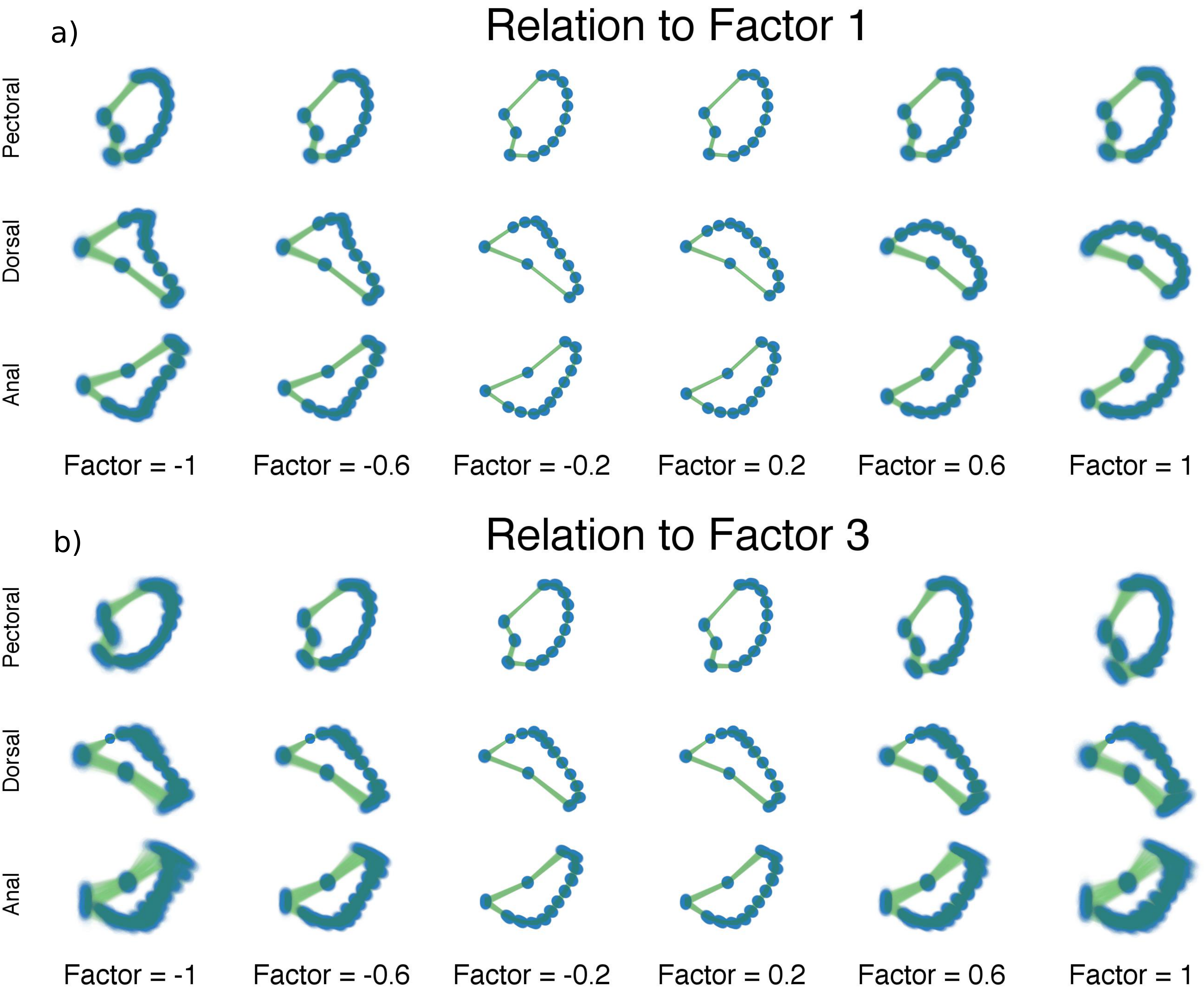}
  \caption{Expected triggerfish fin shape given a range of a) first factor values $\facmat_1$ and b) third factor values $\facmat_3$, holding all others constant. 
Purple dots estimate semi-landmark locations. 
Green lines are interpolated to present a clearer outline of the fin shape. 
For the relation represented by $\facmat_1$ the dorsal and anal fins go from more pointed to less pointed. For the relation represented by $\facmat_3,$ we see a rotation in the pectoral fin.
  }
\label{fig:triggersShapes}
\end{figure}

For 2 of the 5 factors in the $\nfac = 5$ model, Figure \ref{fig:triggersShapes} demonstrates how fin shape changes as a function of latent factor values.
We vary $\facmat_1$ and $\facmat_3$ between $-1$ and $1$ that approximates their highest posterior density range over their reconstructed evolutionary history.
%
%
For $\facmat_1$,
increasing values lead to dorsal and anal fins that become less pointed and more rounded.
For $\facmat_3$,
increasing values lead to a counterclockwise rotation of the dorsal fin.
Our credible band decreases in size as the factor value gets closer to 0 since the standard deviation of the posterior inference on our loadings is multiplied by these factor values as well.

\begin{figure}
    \includegraphics[width=\textwidth]{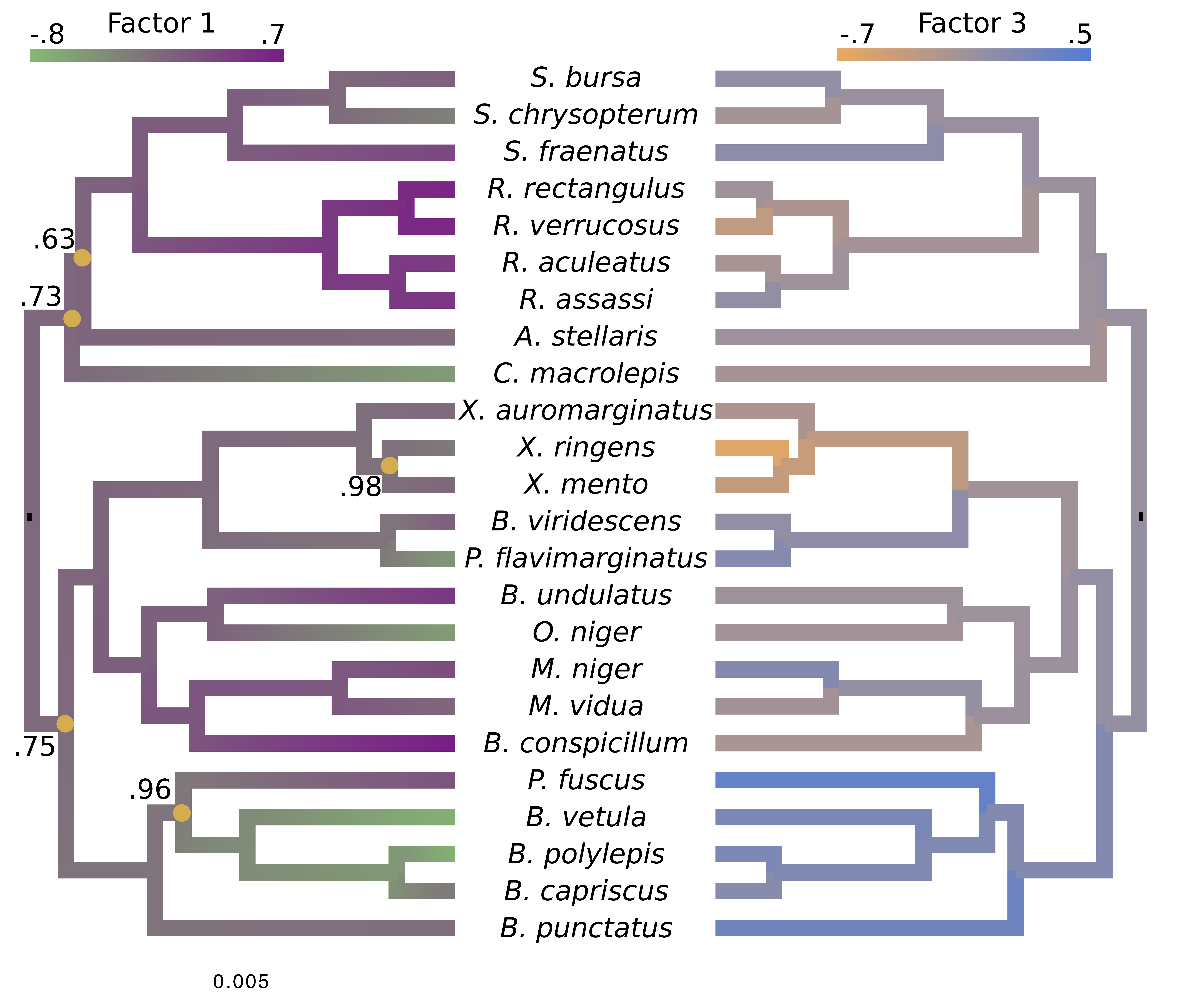}
    \caption{
    Evolution of independent factors $\facmat$ driving triggerfish fin morphology along inferred phylogeny.  
    The colorings display contemporary and ancestral first $\facmat_1$ and third $\facmat_3$ factor values under a $\nfac = 5$ factor PFA model. 
    For $\facmat_1$, green represents positive values and purple represents negative values.
    For $\facmat_3$, the scale is orange to blue.
    The Supplementary Material contains plots for $\facmat_2$, $\facmat_4$ and $\facmat_5$.
    \textit{Balistes polylepis} and \textit{Balistes vetula}
    have negative factor values for the first factor $\facmat_1,$ whereas the clade containing genus \textit{Rhinecanthus} has positive factor values. In the third factor $\facmat_3$, the \textit{Balistes} genus and the species \textit{Pseudobalistes fuscus} have positive factor values whereas the genus \textit{Rhinecanthus} has near 0 factor values. Conversely, the genus \textit{Xanthichthys} has a negative factor value for $\facmat_3$, and has a near 0 value for $\facmat_1$.
    We display the posterior clade probabilities for probabilities $<99\%.$
    }
\label{fig:factree1}
\end{figure}

We also include the corresponding maximum clade credibility (MCC) tree, colored by factor value, with purple representing positive values and green representing negative values for the first factor $\facmat_1,$ and the blue representing positive factor values and orange representing negative factor values for $\facmat_3$ in figure \ref{fig:factree1}.
This tree shows us that the species \textit{Balistes polylepsis} and \textit{Balistes vetula}, have negative factor values for $\facmat_1$, but those species as well as the rest of the clade with the genus \textit{Balistes} and species \textit{Pseudobalistes fuscus} have positive factor values for $\facmat_3,$ whereas the clade containing the genus \textit{Rhinecanthus} has negative factor values for $\facmat_1,$ but a close to 0 factor value for $\facmat_3$. Conversely, the genus \textit{Xanthichthys} has a negative factor value for $\facmat_3,$ and a closer to 0 factor value for $\facmat_1.$ We also display posterior clade probabilities for those clades with probability $<99\%.$

For brevity, we have only considered two factors in this section. We selected $\facmat_1$ and $\facmat_3$ since these factors relate distinctive information, however we include the results for the remaining factors in the supplementary information. 
We additionally include our inference on the precision elements as well as our results on the inference on the other aspects of our tree model in the supplementary material. 

\begin{figure}
\centering
    \includegraphics[width=\textwidth]{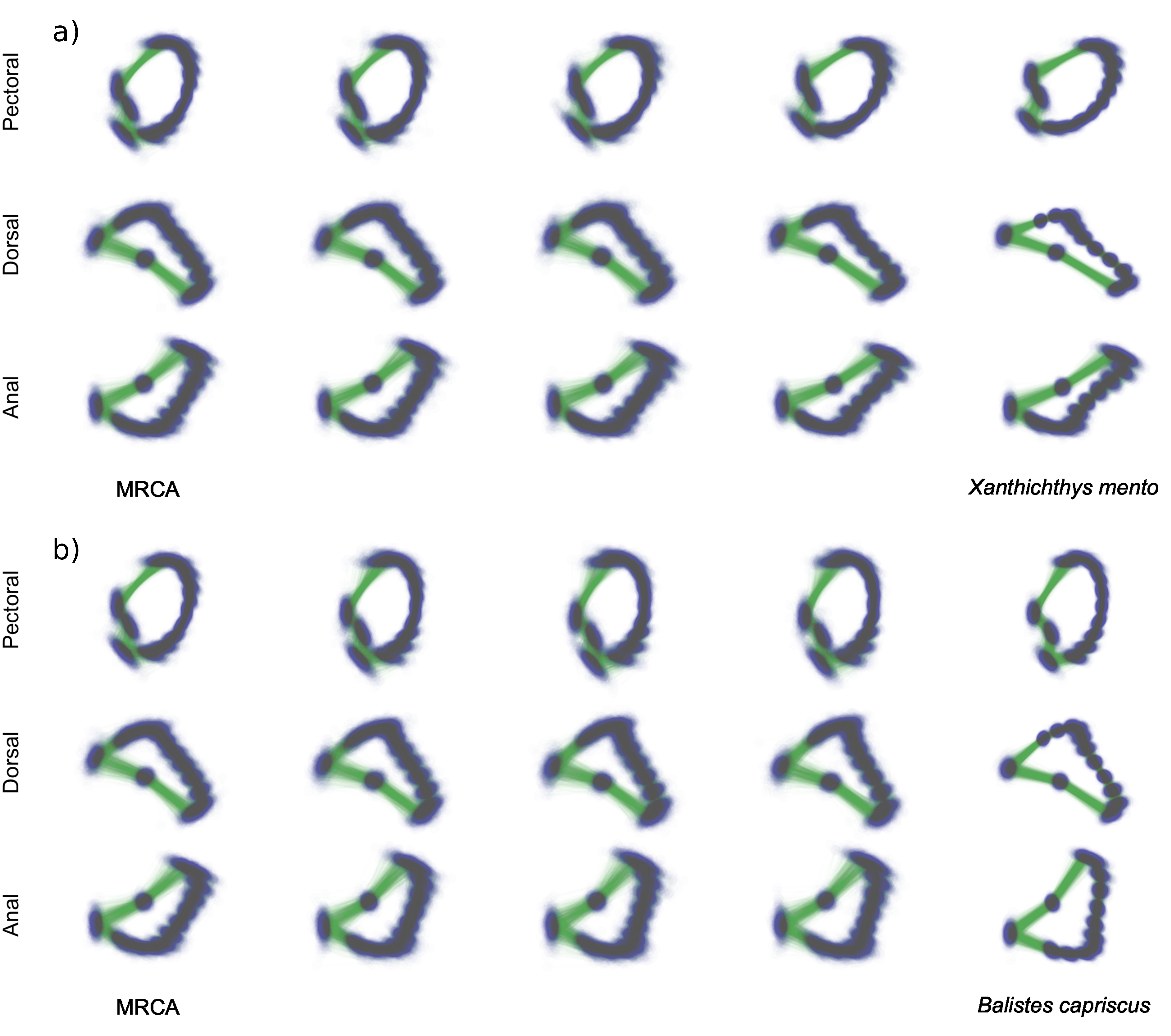}  
      \caption{Inferred ancestral fin shapes at the most recent common ancestor (MRCA) and $1/4$, $1/2$ and $3/4$ of the expected substitution distance between the MRCA and two contemporaneous triggerfish species.
In a), \textit{Xanthichthys mento} has a flat dorsal and anal fin with a point, and a clockwise rotated pectoral fin relative to its ancestors. The dorsal and anal fins become rounder and the pectoral fin rotates counterclockwise moving backwards in time. 
In contrast, in b), \textit{Balistes capriscus} has a broad pointed dorsal and anal fin, and a counterclockwise anal fin. 
 The dorsal and anal fins become more pointed and then round out, while the pectoral fin rotates clockwise.}
\label{fig:triggersAncestral}
\end{figure}

Lastly, PFA facilitates ancestral shape reconstruction.  
Figure \ref{fig:triggersAncestral} depicts inferred pectoral, dorsal and anal fin shapes for ancestors of \textit{Xanthichthys mento} and \textit{Balistes capriscus} at arbitrary points into their evolutionary past.
We choose reconstructions at the most recent 
common 
ancestor (MRCA) of all 24 species in our study and $1/4$, $1/2$ and $3/4$ of the expected sequence substitution distance between the MRCA and both contemporaneous species.
Typically, high aspect ratio fins, or long fins with a small area, are associated with swimming quickly over large distances.
The diet of \textit{Xanthichthys mento} consists mostly of plankton and swims above reefs and has a high aspect ratio, perhaps reflecting a need to hunt down more evasive prey.
We see that these low aspect ratio dorsal and anal fins arose from a moderate MRCA which flatten as the species evolved.
The pectoral fin rotated clockwise as this species evolved.
By contrast,
\textit{Balistes capriscus} has low aspect ratio dorsal and anal fins, reflecting the fact that it swims more towards the reef floors which may be more useful in navigating the complex habitat.
This species evolved from a species with a moderate aspect ratio in its dorsal and anal fins which became broader and more pointed as it evolved.
However, the aspect ratio increases again about $3/4$ of the way through its evolution.
The pectoral fin rotated counterclockwise as it evolved.

This ancestral reconstruction can provide new insights into the trajectories of shape change that could be further investigated with biomechanical and fluid dynamic models.

\section{Computational Aspects}

To draw posterior inference, we simulate MCMC chains of between 200M and 1B steps, subsampling every 10K steps to eliminate unnecessary overhead and ensure the rate-limiting computation remains the PFA and L/MDB transition kernels.  
For path sampling, we employ 100 path points based on the quantiles of a beta $\beta \left( 0.3 , 1 \right)$ random variable \citep{xie2010improving}, with warm-started chains of 10M steps at each point.
In our examples, the PFA chains generate draws three- to five-fold faster than the L/MBD chains.
Further, with the relatively large ratio of latent to non-latent traits in the \textit{Aquilegia} example, we find an approximately 27-fold larger median effective sample size (ESS) across $\loadmat$, $\facmat$ and $\gamma$ than in the latent components of $\boldsymbol{\Sigma}$, demonstrating both faster and more efficient sampling.

\section{Discussion}

This paper merges traditional factor analysis with phylogenetics to provide a new inference tool for comparative studies.
The key connection rests on modeling each factor independently as a Brownian diffusion along a phylogeny.
The tool we provide not only serves as a dimension reduction technique in the face of high-dimensional traits, but directly addresses the principal scientific questions that many comparative studies raise -- specifically, how many independent evolutionary processes are driving these traits?  
Set in a Bayesian framework, we succeed in inferring these processes for combinations of discrete and continuous traits through model selection, while simultaneously accounting for missing measurements and possible phylogenetic uncertainty.

\par

To make inference under PFA practical, we develop two new MCMC integration techniques. While we rely on previously proposed Gibbs samplers for integrating the loading matrix $\loadmat$ and residual trait precisions  $\rowvarresid$, we require an original algorithm based on dynamic programming to integrate the factors $\facmat$ along the phylogeny efficiently.  
Second, we extend path sampling through a softening threshold to handle discrete traits, in which their latent support depends on the path location $\pathParameter$.
Such changing support previously has limited marginal likelihood estimation across many Bayesian models with latent random variables to combine discrete and continuous observations.


\par

In examples involving columbine flower and fish families \textit{Poeciliidae} and \textit{Balistidae} evolution, 
inference
under the PFA is notably quicker under the presence of latent traits, more interpretable and consistently favored via model selection over competing LMBD / MBD models.
%
%
%
Interestingly, this success even holds in the \textit{Poeciliidae} example, where one might expect an LMBD model to outperform.  
Here, the number of parameters inferred in the variance matrix is small relative to the number of parameters that form a PFA.
The \textit{Poeciliidae} and \textit{Balistidae} examples also demonstrate our Bayesian approach's ability to integrate missing data if we make a simple missing-at-random assumption.
%

Unlike many univariate comparative methods, the PFA simultaneously adjusts for correlation between all traits.
This advantage reveals that some previously identified trait relationships in \textit{Poeciliidae} evolution may be spurious.
Further, as demonstrated in the columbine flower example, the inferred factors and their associated loadings probabilistically cluster traits into independent processes that provide additional scientific insight, often hard to discern from the correlation matrix that a LMBD model provides.  
%

\par

An important computational limitation of PFA arises when the number of taxa $\ntaxa$ is much greater than the number of traits $\ntraits$.
For the PFA, computational cost of our current MCMC integration scales as 
$\order{\ntaxa^2 \nfac + \ntaxa\nfac^2\ntraits}$, while the cost is $\order{\ntaxa \ntraits^2}$ for the LMBD / MBD models.
Nonetheless,
the \textit{Poeciliidae} example carries $\ntaxa / \ntraits \approx 9$ and, still, the PFA model integrates about $3 \times$ more efficiently due to the example's large ratio of latent traits.
For larger $\ntaxa / \ntraits$ ratios, we are currently devising algorithms that remain linear in $\ntaxa$ as future work. 

Arguably,
PFA reaches its greatest potential when the number of traits stands large relative to the number of taxa -- the reputed ``large $\ntraits$, small $\ntaxa$'' setting.
This setting arises commonly in the field of geometric morphometrics where very long series of Cartesian, (semi-) land-mark coordinate measurements define the shape of the organism.
In our \textit{Balistidae} example, the PFA identifies a number of independent evolutionary processes driving pectoral, dorsal and anal fin shapes.
With the help of sequence data, the PFA also simultaneously infers the phylogeny and reconstructs ancestral shapes.
We believe that morphometrics stands poised as a prime beneficiary of PFA. 

One potential extension of this method comes from \cite{brownian_diffusion_phylogeography}, where they place different diffusion rates on different branches. 
Additionally we can adapt the methods in \cite{GillDrift}, which allows us to incorporate inference on drift in our factors whose direction changes at different points in the evolutionary process.
Both of these methods are implemented in BEAST and are therefore easily adapted.
%
%



\section{Acknowledgments}

The research leading to these results has received funding from the European Research Council under the European Community's Seventh Framework Programme (FP7/2007-2013) under
Grant Agreement no.~278433-PREDEMICS and ERC Grant agreement no.~260864 and the
National Institutes of Health (R01 AI107034,  R01 AI117011 and R01 HG006139) and the
National Science Foundation (DMS 1264153).


\bibliography{bibliography}

\begin{appendices}

\renewcommand{\thesubsubsection}{\Alph{subsubsection}}
\newcommand\pcdensity[3]{p^{#3}\left(#1 | #2 \right)}
\newcommand\tab{\hspace{.5in}}

\subsubsection{Phylogenetic factor analysis Gibbs sampling}
While the Gibbs samplers for a standard factor analysis are known and well documented \citep{LopesWest}, there are two aspects of our phylogenetic model that differ sufficiently to require a fresh look at how to draw posterior inference. 
First, our prior on $\facmat$ is based on a phylogenetic tree and therefore requires particular consideration in order to produce an efficient Gibbs sampler. 
Second, our inference on $\nfac$ uses a path sampling approach where we need to infer $\loadmat,$ $\facmat,$ and $\rowvarresid$ at each point along the path $\modPath{\pathParameter, \datamat, \sequence, \paramList},$  and deriving a Gibbs sampler that works for any point in the path $\pathParameter$ will aid this process. 

\paragraph{Sampling factors}
\label{app:gibbs}
In a standard Bayesian factor analysis, the prior on each element $\facmatsm_{\ida\idb}$ is $N(0,1),$ and so the entire matrix \(\facmat\) can be Gibbs sampled efficiently in a single step \citep{LopesWest}. For the phylogenetic factor analysis model, the prior on the factors is defined by Brownian motion on a phylogenetic tree as defined in (\ref{eq:factorModel}). Thus the conditional density of $\facmat | \workmat, \loadmat, \rowvarresid$ in our model is proportional to
\begin{equation}
\begin{aligned}
\cdensity{\workmat}{\facmat, \loadmat, \rowvarresid} \density{\facmat}
&\propto \exp  \left\{-\frac{1}{2}\trOperator{
	\left(\workmat - \facmat\loadmat \right)
	 \rowvarresid
	 \left(\workmat - \facmat\loadmat \right)\transpose}\right\} \\ 
	  &\tab\times  \exp \left\{-\frac{1}{2}\trOperator{\facmat\facmat\transpose \left(\facprecTree  + \pss^{-1} \mathbf{J}\right)^{-1}}\right\}.
	 \end{aligned}
\end{equation}
This expression does not appear to represent a distribution from which we can easily sample, principally stemming from the fact that $\rowvarresid$ is a between-column precision and $\facprecTree  + \pss^{-1} \mathbf{J}$ is a between-row precision.

Fortunately, \cite{cybis2015assessing} devise a pre-order tree-traversal algorithm to determine the conditional distribution $\facmat_{\ida.}\transpose | \facmat_{-\ida .}$ of the factors at a single tip given all other tip values.  This distribution is multivariate normal \(\text{MVN}(\facmeancond, \facpreccond)\) with conditional mean \(\facmeancond\) and conditional precision \(\facpreccond\).
%
Further, in order to numerically estimate $\facmat$ at any point along the path $\modPath{\pathParameter, \datamat, \sequence, \paramList},$ we define
\begin{equation}
\modPath{\facmat_{\ida .} | \pathParameter, \deltaIndicator{\ida}\workmat, \facmat_{-\ida.}, \loadmat, \rowvarresid} \propto \jlikelihood{\deltaIndicator{\ida}\workmat | \facmat_{\ida.}, \loadmat, \rowvarresid}^\pathParameter \working{\facmat_{\ida.} | \facmat_{-\ida.}}.
\end{equation}
Substituting in the appropriate densities and completing the square, we find that this path is proportional to
\begin{equation}
\begin{aligned}
 	q^*\left(\facmat_{\ida .} | \pathParameter, \right. &\deltaIndicator{\ida} \left.\workmat, \facmat_{-\ida.}, \loadmat, \rowvarresid\right)
	\\
	 &\propto  \exp\left\{-\frac{1}{2} \pathParameter 
	\left(\deltaIndicator{\ida}\transpose\workmat - \facmat_{\ida \cdot} \loadmat\right) \rowvarresid
	\left(\deltaIndicator{\ida}\transpose\workmat - \facmat_{\ida \cdot} \loadmat\right)\transpose\right\} \\
	& \tab\times\exp{\left\{-\frac{1}{2}\left(\facmat_{\ida \cdot}\transpose - \facmeancond\right)\transpose\facpreccond\left(\facmat_{\ida \cdot}\transpose - \facmeancond\right)\right\}} \\
	&\propto \exp{\left\{-\frac{1}{2} \facmat_{\ida .}\left(\pathParameter\loadmat \rowvarresid \loadmat\transpose + \facpreccond\right)
	\right. 
	\left. \facmat_{\ida \cdot}\transpose 
	- 2 \facmat_{\ida \cdot}\left(\pathParameter\loadmat\rowvarresid\workmat\transpose\deltaIndicator{\ida} 
	+ \facpreccond\facmeancond\right)\vphantom{\frac{1}{2}}\right\}}\\
\label{eq:MVNkernel}
&\propto \exp\left\{-\frac{1}{2}
\left(\facmat_{\ida .}\transpose - \gibbsMeanBeta{\facmat}{\ida}\right)\transpose
\left( \gibbsVarianceBeta{\facmat}{\ida} \right)^{-1}
 \left(\facmat_{\ida .}\transpose - \gibbsMeanBeta{\facmat}{\ida}\right)
 \right\},
\end{aligned}
\end{equation}
where
\begin{align}
\gibbsMeanBeta{\facmat}{\ida} &= \gibbsVarianceBeta{\facmat}{\ida}
	\left(
		\pathParameter
		\loadmat
		\rowvarresid
		\workmat\transpose
		\deltaIndicator{\ida}
		+ 
		\facpreccond\facmeancond 
	\right) \label{eq:factMean2} \\
\intertext{and}
\gibbsVarianceBeta{\facmat}{\ida} &= 				
	\left(
		\pathParameter\loadmat\rowvarresid\loadmat\transpose + \facpreccond
	\right)^{-1} . \label{eq:factVar2}
\end{align}
Equation (\ref{eq:MVNkernel}) is proportional to the density of a $\text{MVN}\left(\gibbsMeanBeta{\facmat}{\ida}, \gibbsVarianceBeta{\facmat}{\ida}\right)$; therefore, in order to  sample $\facmat$ at a particular point in the path $\pathParameter,$ we can draw a row $\facmat_{\ida.}$ from the distribution $\text{MVN}\left(\gibbsMeanBeta{\facmat}{\ida}, \gibbsVarianceBeta{\facmat}{\ida}\right)$.

\paragraph{Sampling loadings}

The loadings matrix can be Gibbs sampled using the same method described by \citet{LopesWest} with an additional adaptation for use in path sampling. For the examples provided in this paper, we place a \(N(0,1)\) prior on each cell in the loadings matrix; however, in this section we prove the Gibbs Sampler for a generic \(N(\loadmean, \loadprec)\) prior. To begin, we again define for a point on the path $\pathParameter,$
\begin{equation}
\modPath{\loadmat | \pathParameter, \workmat, \facmat, \rowvarresid, \loadmean, \loadprec} =\jlikelihood{\workmat | \loadmat, \facmat, \rowvarresid, \loadmean, \loadprec}^{\pathParameter}\working{\loadmat} .
\end{equation}
Plugging in the proper values for the sampling density and priors, rearranging and completing the square, we find that
\begin{equation}
\begin{aligned}
q^* \left(\loadmat | \pathParameter, \right. & \left.\workmat, \facmat, \rowvarresid, \loadmean, \loadprec \right)
\\
&\propto\exp\left\{-\frac{1}{2}\pathParameter\text{tr}
	\left[\left(\workmat-\facmat\loadmat\right)
		\rowvarresid
	\left(\workmat-\facmat\loadmat\right)\transpose\right]\right\}\\
	& \tab \times\exp\left\{-\frac{1}{2}\text{tr}\left[\left(\loadmat-\loadmean\boldsymbol{1}\right)
\loadprec\boldsymbol{I}
\left(\loadmat-\loadmean\boldsymbol{1}\right)
\transpose\right] \right\}
\\ 
&\propto \exp\left\{-\frac{1}{2}\text{tr}\left[\pathParameter\facmat\loadmat\rowvarresid\loadmat\transpose\facmat\transpose
-2\pathParameter\workmat\rowvarresid\loadmat\transpose
\facmat\transpose+\loadprec\loadmat\loadmat\transpose
-2\loadprec\loadmean\boldsymbol{1}\loadmat\transpose\right] \right\}
\\&=\exp\left\{-\frac{1}{2}\text{tr}\left[
\pathParameter\loadmat\transpose\facmat\transpose\facmat\loadmat\rowvarresid
+\loadprec\loadmat\transpose\loadmat
-2\left(\pathParameter\rowvarresid\workmat\transpose\facmat\loadmat
+\loadprec\loadmean\boldsymbol{1}\transpose\loadmat\right)\right] \right\}
\\
\label{eq:loaddensity}
&\propto\prod^\ntraits_{\idb = 1}
\exp
\left\{ - \frac{1}{2}
\left(\loadmat_{.\idb}-\gibbsMeanBeta{\loadmat}{\idb}\right)\transpose
\left(\gibbsVarianceBeta{\loadmat}{\idb}\right)^{-1}
\left(\loadmat_{.\idb}-\gibbsMeanBeta{\loadmat}{\idb}\right)
\right\},
\end{aligned}
\end{equation}
 where 
$\loadmat_{.\idb} = 
\left( \loadmatsm_{1\idb}, \ldots, \loadmatsm_{\idp\idb} \right)$, 
$\mathbf{1}$ is a matrix of 1's with the same dimensions as $\loadmat$, 
\begin{align}
\gibbsMeanBeta{\loadmat}{\idb} &=
	 \gibbsVariance{\loadmat}{\idb} \pathParameter\rowvarresidsm_\idb\facmat_{1:\idp}\transpose\workmat \, \deltaIndicator{\idb} 
\intertext{and}
\gibbsVarianceBeta{\loadmat}{\idb} &=
	\left(
		\pathParameter \rowvarresidsm_\idb\facmat_{1:\idp}\transpose\facmat_{1:\idp} + \mathbf{I}_{\idp}
	\right)^{-1}.
\end{align}
Hence we find the expression in (\ref{eq:loaddensity}) is proportional to a product of independent $\text{MVN}\left(
\gibbsMeanBeta{\loadmat}{\idb},
\gibbsVarianceBeta{\loadmat}{\idb}
\right)$ densities.
Therefore, if we wish to numerically sample a loadings column $\loadmat_{.\idb}$ at a point on the path $\pathParameter$ then we can sample from the distribution $\text{MVN}\left(
\gibbsMeanBeta{\loadmat}{\idb},
\gibbsVarianceBeta{\loadmat}{\idb}
\right).$
Since the densities across columns are independent, we may sample from them in parallel.

\paragraph{Sampling residual precision}
\def\cont{c}
We wish to sample $\rowvarresid$ at any point in our path $\modPath{\pathParameter, \datamat, \sequence, \paramList}.$ Let \(\rowvarresid_{\cont}\) be a matrix equivalent to $\rowvarresid$ with rows and columns corresponding to discrete traits removed. We then say that $\rowvarresid_\cont=\left(\rowvarresidsm_{\left(1\right)}, \ldots, \rowvarresid_{\left(\ntraits'\right)}\right)\transpose$ where $\rowvarresidsm_{\left(\idb\right)}$ models continuous trait $\idb$ and $\ntraits'$ is the number of continuous traits in our model. 
If we define $\loadmat_\cont$ and $\workmat_\cont$ as the matrices $\loadmat$ and $\workmat$ with the columns corresponding to discrete traits removed, then we can say $\workmat_\cont \sim \text{MVN}(\facmat\loadmat_\cont, \rowvarresid_\cont).$ 
Our prior on $\rowvarresidsm_{\left(\idb\right)}$ is i.i.d. for different values of $\idb$ and has distribution
$ \Gamma(\rowvarresidshape,\rowvarresidrate).
$
 For an arbitrary point $\pathParameter$ in our path $\modPath{\pathParameter, \datamat, \sequence, \paramList},$ we then define
\begin{equation}
\modPath{\rowvarresid_\cont | \pathParameter, \workmat_\cont, \facmat, \loadmat_\cont} 
\propto \jlikelihood{\workmat_\cont | \rowvarresid_\cont, \facmat, \loadmat_\cont}^{\pathParameter} 
\working{\rowvarresid_\cont},
\end{equation} 
with density
\begin{equation}
\begin{aligned}
q^* \left(\rowvarresid_\cont | \pathParameter,\right. &\left.\workmat_\cont, \facmat, \loadmat_\cont\right) \\ 
 &\propto 
 \prod_{\idb=1}^{\ntraits'} \rowvarresidsm^{\pathParameter\ntaxa/2}_{\left(\idb\right)} 
\times 
\exp\left\{-\frac{\pathParameter}{2}\left[\deltaIndicator{\idb}\transpose\left(\workmat_\cont-\facmat\loadmat_\cont\right)\transpose 
\left(\workmat_\cont-\facmat\loadmat_\cont\right)
\deltaIndicator{\idb}\rowvarresidsm_{\left(\idb\right)}\right]\right\}
 \\ &\tab \times
\prod_{\idb=1}^{\ntraits'} \rowvarresidsm_{\left(\idb\right)}^{\rowvarresidshape-1} \times\exp\left\{- \rowvarresidrate\rowvarresidsm_{\left(\idb\right)}\right\}
\\
\label{eq:gamRes}
&= \prod_i^{\ntraits'} \rowvarresidsm_{\left(\idb\right)}^{\rowvarresidshape+\pathParameter\ntaxa/2-1}
\times
\exp\left\{-\left(
\rowvarresidrate + \frac{\pathParameter}{2}\deltaIndicator{\idb} \transpose
\left(\workmat_\cont-\facmat\loadmat_\cont\right)\transpose
\left(\workmat_\cont-\facmat\loadmat_\cont\right)\deltaIndicator{\idb} 
\right)\rowvarresidsm_{\left(\idb\right)} \right\}.
\end{aligned}
\end{equation}
The expression in (\ref{eq:gamRes}) is proportional to the density of a gamma $\gammaDistribution{
	\rowvarresidshape + \frac{\pathParameter\ntaxa}{2}
}{
\rowvarresidrate + \frac{\pathParameter}{2} \,
	\deltaIndicator{\idb}\transpose
	\left( 			
		\workmat - \facmat \loadmat
	\right)\transpose
	\left(
		\workmat - \facmat \loadmat
	\right)
	\deltaIndicator{\idb}
}$ random variable,
and therefore we can sample from this gamma distribution in order to sample $\rowvarresidsm_{\left(\idb\right)}$ at a given point in the path.

\end{appendices}
\end{document}


\begin{flushright}
Version dated: \today
\end{flushright}

\bigskip
\medskip

{\Large \noindent \sc Supplementary material for:}
\begin{center}

\noindent{\Large \bf Phylogenetic Factor Analysis}
\bigskip

\noindent{\normalsize \sc
	Max R.~Tolkoff$^1$,
	Michael E.~Alfaro$^{2}$,
	Guy Baele$^{3}$,
	Philippe Lemey$^{3}$, \\
    and Marc A.~Suchard$^{1,4,5}$}\\
\noindent {\small
  \it $^1$Department of Biostatistics, Jonathan and Karin Fielding School of Public Health, University
  of California, Los Angeles, United States} \\
  \it $^2$Department of Ecology and Evolutionary Biology, University of California, Los Angeles, United States \\
  \it $^3$Department of Microbiology and Immunology, Rega Institute, KU Leuven, Leuven, Belgium \\
  \it $^4$Department of Biomathematics, David Geffen School of Medicine at UCLA, University of California,
  Los Angeles, United States \\
  \it $^5$Department of Human Genetics, David Geffen School of Medicine at UCLA, Universtiy of California,
  Los Angeles, United States

\end{center}
\medskip
\noindent{\bf Corresponding author:} Marc A. Suchard, Departments of Biostatistics, Biomathematics, and Human Genetics, 
University of California, Los Angeles, 695 Charles E. Young Dr., South,
Los Angeles, CA 90095-7088, USA; E-mail: \url{msuchard@ucla.edu}\\

\vspace{1in}

\renewcommand\thesection{S\arabic{section}}
\renewcommand\thesubsubsection{S\arabic{subsubsection}}
\renewcommand\thefigure{S\arabic{figure}}
\renewcommand\thetable{S\arabic{table}}

\newpage
We perform a phylogenetic factor analysis (PFA) on $\ntaxa=24$ triggerfish species with 13 $\left(x, y\right)$ coordinate measurements on the pectoral, dorsal and anal fins $\left(\ntraits=78\right)$ obtained by \cite{triggerfish1}. We additionally use 12S (833 nucleotides, nt), and 16S (563 nt) mitochondrial genes and RAG1 (1471 nt), rhodopsin (564 nt) and Tmo4C4 (575 nt) nuclear genes obtained by \cite{triggerfish2} with a Kingman coalescent prior on the tree topology \citep{kingman_coalescent}, an HKY substitution model \citep{hasegawa85} as well as a discretized, one-parameter Gamma distribution with unknown shape and proportion of invariant sites \citep{gammaSiteRate}. We settle on a $\nfac=5$ factor model.
\subsubsection{Triggerfish Fin Precision Elements}

\begin{longtable}{clD{.}{.}{-1}D{.}{.}{8}D{.}{.}{-1}D{.}{.}{8}cccccc}
\caption{Triggerfish pectoral, dorsal and anal fin precision element posterior mean (mean) and 95\% Bayesian credible interval (BCI) estimates.}
\cr
&\multicolumn{1}{c}{Label} 
&\multicolumn{1}{c}{X-mean}  
&\multicolumn{1}{c}{X-95\% BCI}
&\multicolumn{1}{c}{Y-mean}  
&\multicolumn{1}{c}{Y-95\% BCI} \\
\hline
 & Pectoral Pt.~1 & 12.90 & [5.27, 21.95] & 1.13 & [0.46, ~1.83] \\ 
 & Pectoral Pt.~2 & 14.29 & [5.72, 24.25] & 2.29 & [0.99, ~3.80] \\ 
 & Pectoral Pt.~3 & 19.34 & [7.53, 32.08] & 7.14 & [2.80, 11.97] \\ 
 & Pectoral Pt.~4 & 12.27 & [4.97, 20.30] & 10.84 & [3.90, 18.14] \\ 
 & Pectoral Pt.~5 & 2.84 & [1.14, ~4.64] & 14.86 & [5.93, 25.05] \\ 
 & Pectoral Pt.~6 & 0.95 & [0.43, ~1.55] & 13.43 & [5.56, 22.71] \\ 
 & Pectoral Pt.~7 & 2.43 & [1.01, ~4.04] & 8.86 & [3.39, 15.41] \\ 
 & Pectoral Pt.~8 & 9.52 & [3.80, 16.06] & 3.93 & [1.37, ~7.00] \\ 
 & Pectoral Pt.~9 & 15.20 & [6.36, 26.11] & 1.80 & [0.63, ~3.13] \\ 
 & Pectoral Pt.~10 & 12.07 & [4.53, 20.23] & 4.92 & [1.96, ~8.47] \\ 
 & Pectoral Pt.~11 & 6.07 & [2.62, 10.24] & 11.00 & [4.87, 18.79] \\ 
 & Pectoral Pt.~12 & 2.75 & [1.11, ~4.51] & 6.22 & [2.55, 10.50] \\ 
 & Pectoral Pt.~13 & 1.09 & [0.49, ~1.85] & 10.86 & [4.33, 18.27] \\ 
 \hline
 & Dorsal Pt.~1 & 12.56 & [4.82, 21.75] & 6.55 & [1.89, 12.08] \\ 
 & Dorsal Pt.~2 & 11.26 & [3.83, 18.93] & 7.30 & [2.48, 12.71] \\ 
 & Dorsal Pt.~3 & 10.89 & [3.88, 18.53] & 3.69 & [1.43, ~6.05] \\ 
 & Dorsal Pt.~4 & 3.64 & [1.40, ~6.20] & 2.83 & [1.22, ~4.70] \\ 
 & Dorsal Pt.~5 & 2.18 & [0.80, ~3.82] & 2.46 & [1.02, ~4.11] \\ 
 & Dorsal Pt.~6 & 6.38 & [2.01, 11.38] & 3.24 & [1.20, ~5.75] \\ 
 & Dorsal Pt.~7 & 14.76 & [5.16, 25.55] & 7.55 & [2.19, 13.70] \\ 
 & Dorsal Pt.~8 & 13.62 & [5.09, 22.87] & 5.33 & [1.53, 10.30] \\ 
 & Dorsal Pt.~9 & 12.12 & [4.19, 21.13] & 2.89 & [1.04, ~5.02] \\ 
 & Dorsal Pt.~10 & 8.62 & [2.19, 16.12] & 3.15 & [1.24, ~5.26] \\ 
 & Dorsal Pt.~11 & 5.21 & [1.50, ~9.91] & 3.70 & [1.44, ~6.16] \\ 
 & Dorsal Pt.~12 & 2.43 & [0.95, ~4.03] & 3.86 & [1.55, ~6.38] \\ 
 & Dorsal Pt.~13 & 1.99 & [0.81, ~3.33] & 3.37 & [1.33, ~5.80] \\ 
 \hline
 & Anal Pt.~1 & 6.32 & [2.44, 10.71] & 8.15 & [2.86, 14.14] \\ 
 & Anal Pt.~2 & 8.77 & [3.46, 15.15] & 7.40 & [2.87, 13.12] \\ 
 & Anal Pt.~3 & 10.49 & [3.91, 17.73] & 2.28 & [0.90, ~3.74] \\ 
 & Anal Pt.~4 & 11.81 & [4.37, 20.06] & 1.70 & [0.74, ~2.85] \\ 
 & Anal Pt.~5 & 4.79 & [1.67, ~8.26] & 3.24 & [1.22, ~5.57] \\ 
 & Anal Pt.~6 & 3.01 & [1.04, ~5.11] & 4.04 & [1.36, ~7.08] \\ 
 & Anal Pt.~7 & 4.34 & [1.77, ~7.54] & 6.13 & [2.00, 11.02] \\ 
 & Anal Pt.~8 & 6.69 & [2.56, 11.28] & 9.65 & [3.27, 16.94] \\ 
 & Anal Pt.~9 & 14.89 & [5.50, 25.09] & 9.95 & [3.71, 17.29] \\ 
 & Anal Pt.~10 & 15.39 & [6.22, 26.70] & 7.76 & [2.88, 13.26] \\ 
 & Anal Pt.~11 & 1.40 & [0.58, ~2.35] & 4.45 & [1.68, ~7.52] \\ 
 & Anal Pt.~12 & 4.20 & [1.71, ~7.04] & 3.24 & [1.22, ~5.46] \\ 
 & Anal Pt.~13 & 8.29 & [3.12, 14.11] & 5.50 & [2.15, ~9.30] \\ 
 \end{longtable}


\clearpage
\subsubsection{Remaining loadings plots for triggerfish example}
    \begin{figure}[!htb]
    \center
    \includegraphics[width=.81\textwidth]{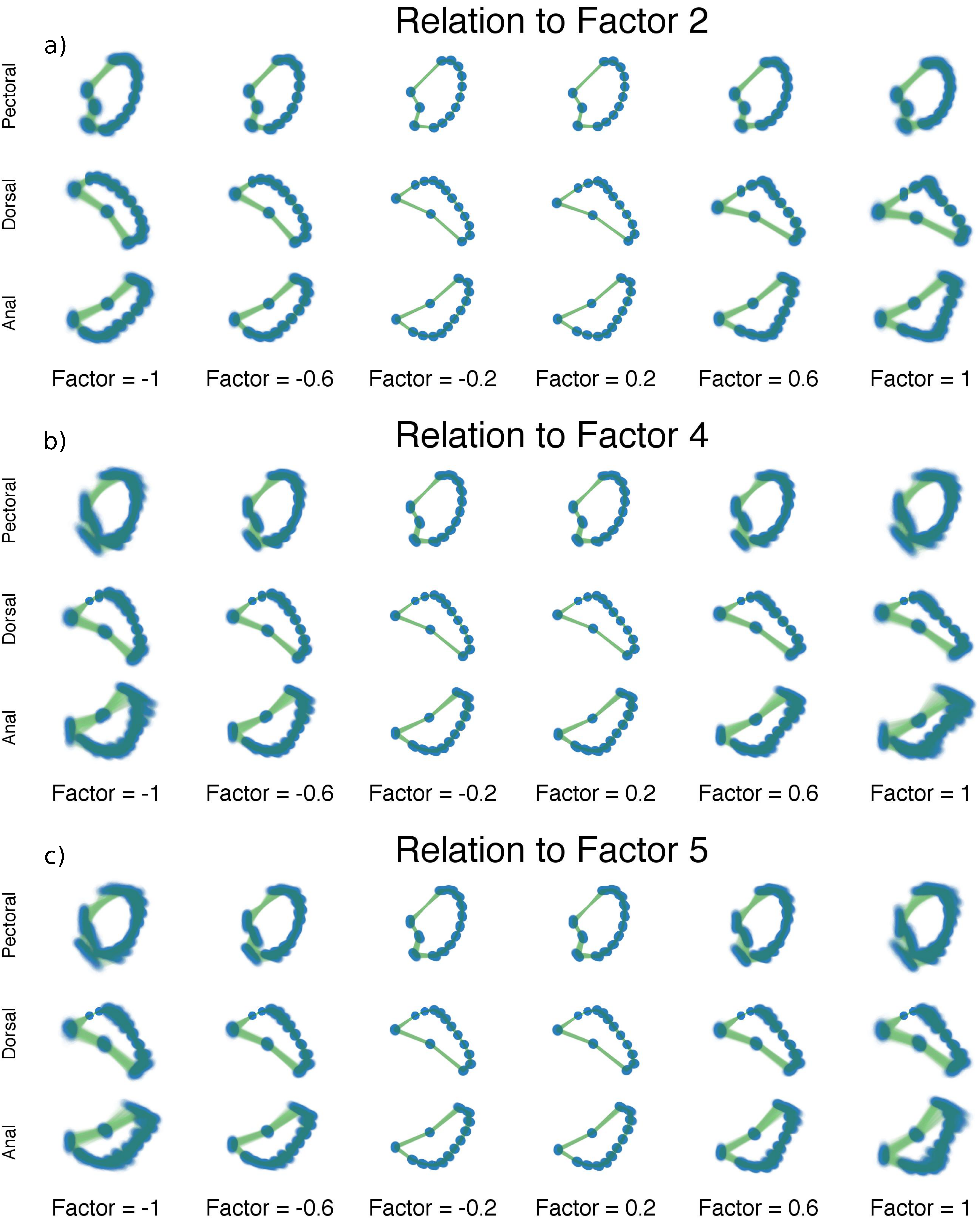}
    \caption{Expected triggerfish fin shape given a range of a) $\facmat_2$, b) $\facmat_4$ and c) $\facmat_5$ values, holding other factor values constant.     
    Purple dots estimate semi-landmark locations. 
    Green lines are interpolated to present a clearer outline of the fin shape.}
\end{figure}
\label{sec:tfacload}
\clearpage
\subsubsection{Remaining factor tree plots for triggerfish example}
    \begin{figure}[!htb]
    \includegraphics[width=.9\textwidth]{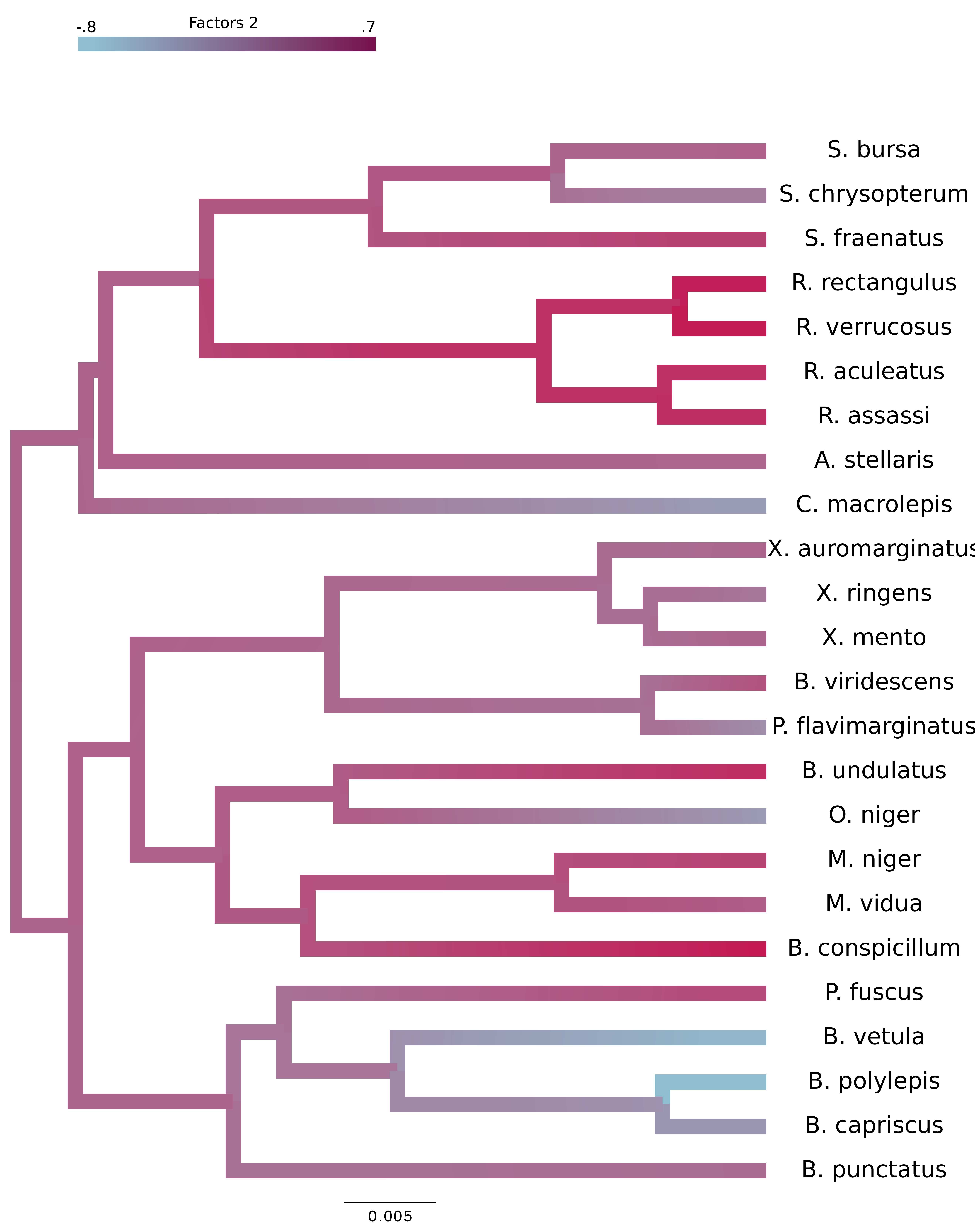}
        \caption{Maximum clade credibility tree for triggerfish species with teal representing a negative factor value and red-purple representing larger factor values.}
    \end{figure}
    \begin{figure}[!htb]
    \includegraphics[width=.9\textwidth]{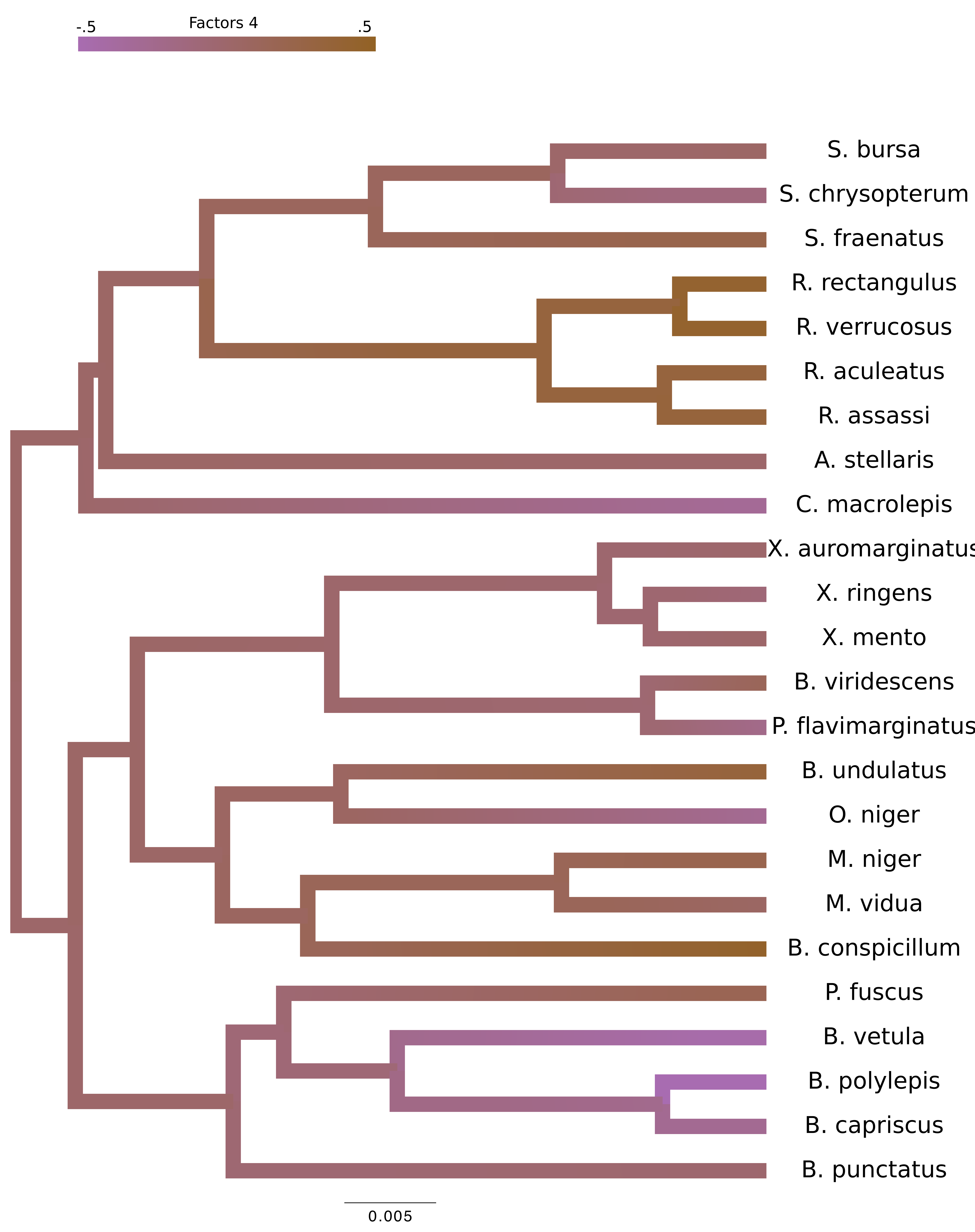}
        \caption{Maximum clade credibility tree for triggerfish species with light purple representing a negative factor value and brown representing larger factor values.}
        \end{figure}
    \begin{figure}[!htb]
    \includegraphics[width=.9\textwidth]{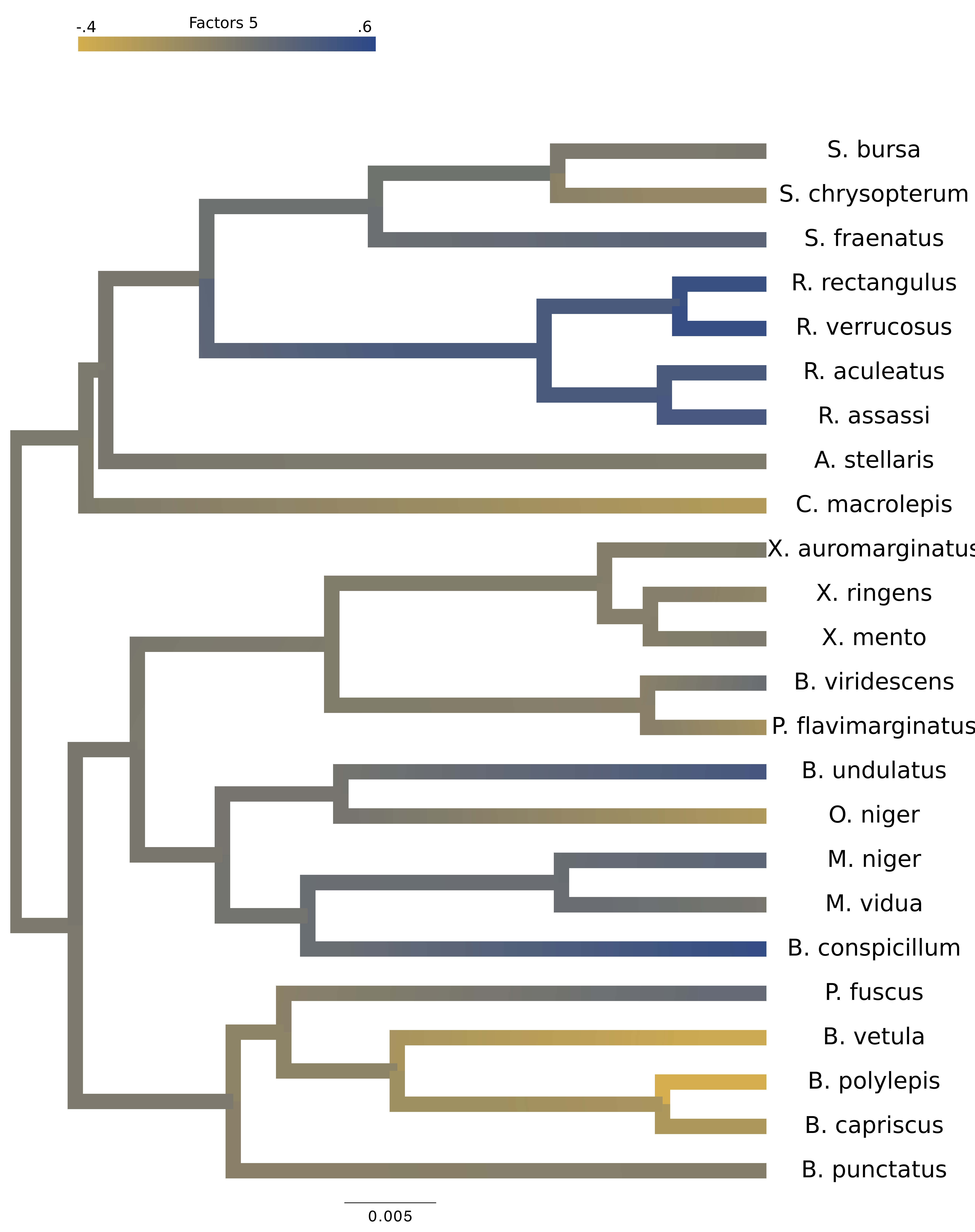}
    \caption{Maximum clade credibility tree for triggerfish species with yellow representing a negative factor value and navy blue representing larger factor values.}
\end{figure}
\label{sec:tfactree}

\clearpage
\subsubsection{Phylogenetic character substitution estimates}
\begin{table}[!htb]
\caption{Posterior estimates of HKY substitution model \citep{hasegawa85}, discretized Gamma shape $\alpha,$ and proportion of invariant sites $P_{inv}$ \citep{gammaSiteRate}. For the HKY model, $\left(\pi_A, \pi_C, \pi_G, \pi_T\right)$ represent the nucleotide stationary distribution, and $\kappa$ represents the rate ratio of transitions to transversions.}
\begin{center}
\begin{tabular}{crD{.}{.}{-1}D{.}{.}{-1}}
& 
 &
\multicolumn{1}{c}{Posterior}  &
\multicolumn{1}{c}{\hspace{.4in}95\% Bayesian }
\\ & \multicolumn{1}{c}{Trait} & \multicolumn{1}{c}{mean}&  \multicolumn{1}{c}{\hspace{.4in}credible interval}
\\
 \hline
 & $\pi_A$ & 0.275 & [0.262, 0.288] \\
 & $\pi_C$ & 0.259 & [0.247, 0.271] \\
 & $\pi_G$ & 0.221 & [0.231, 0.231] \\
 & $\pi_T$ & 0.245 & [0.232, 0.256] \\
 & $\kappa$ & 4.304 & [3.852, 4.816]\\
 & $\alpha$ & 0.552 & [0.382, 0.753]  \\
 & $P_{inv}$ & 0.673 & [0.627, 0.725] 
 \end{tabular}
 \end{center}
\end{table}
\newpage
\bibliography{bibliography}